\documentclass[prl,aps,amssymb,amsmath]{article}
\usepackage{amssymb,amsmath,tikz}
\usepackage{amsthm}
\usepackage{mathrsfs}
\usepackage{pifont}

\showoutput
\DeclareMathOperator*{\p}{p}

\newcommand{\ud}{\mathrm{d}}

\theoremstyle{plain}

\newtheorem*{twr*}{THEOREM}
\newtheorem*{lem*}{LEMMA}

\newtheorem*{rem*}{REMARK}

\newtheorem*{notn*}{NOTATION}
\newtheorem*{wiener-ito*}{WIENER-IT\^O-SEGAL DECOMPOSITION}

\begin{document}
\title{{\bf On single-photon wave function}}
\author{Jaros{\l}aw Wawrzycki \footnote{Electronic address: jaroslaw.wawrzycki@wp.pl or jaroslaw.wawrzycki@ifj.edu.pl}
\\Institute of Nuclear Physics of PAS, ul. Radzikowskiego 152, 
\\31-342 Krak\'ow, Poland}
\maketitle

\vspace{1cm}

\begin{abstract}
We present in this paper how the single-photon wave function for transversal photons (with the direct sum of 
ordinary unitary representations of helicity 1 and -1
acting on it) is subsumed within the formalism of Gupta-Bleuler for the quantized free electromagnetic field in the Krein space (i.e. in the ordinary Hilbert space endowed with the Gupta-Bleuler operator $\eta$). 
Rigorous Gupta-Bleuler quantization of the free electromagnetic field 
is based on a generalization of ours (published formerly) of the Mackey theory of induced representations which includes representations preserving the indefinite Krein inner-product given by the Gupta-Bleuler operator and acting in the Krein space. The free electromagnetic field is constructed by application of the direct sum of (symmetrized) tensor products of a specific indecomposable (but reducible) single-photon representation which is Krein-isometric but non unitary, we call it {\L}opusza\'nski representation, i. e. we construct the field by application of the Segal's second quantization functor to the specific Krein-isometric representation. A closed subspace
$\mathcal{H}_{\textrm{tr}}$ of the single-photon Krein space on which the indefinite Krein-inner-product is strictly positive
is constructed such that the Krein-isometric single-photon representation generates 
modulo unphysical states precisely the action of a representation which preserves the positive inner product
on $\mathcal{H}_{\textrm{tr}}$ induced by the Krein inner product,
and is equal to the direct sum of ordinary unitary representations of helicity 1 and -1 respectively.  Two states of single photon Krein space are physically equivalent whenever differ by a state of Krein norm zero and whose projection
 on $\mathcal{H}_{\textrm{tr}}$, in the sense of the Krein-inner-product, vanishes. In particulart it follows that the results of Bia{\l}ynicki-Birula on the single-photon wave function may be reconciled with the micro-local perturbative approach to QED initiated by St\"uckelberg and Bogoliubov.

\end{abstract}

\section{Introduction}

The wave function of single photon in the position picture is a concept which is accompanied with
controversial opinions. For example in \cite{bohm} or \cite{power} there is even stated that position wave function 
for photon does not exist. We agree e.g. with \cite{Bialynicki} and \cite{bialynicki-2} and the authors cited there,
that quantum field theory speaks for the contrary: generally a free quantum field is constructed by the application of
the symmetrized/antisymmetrized tensoring and direct sum operations (the so called second quantized functor) to a specific representation of the double covering of the Poincar\'e group acting in a space, which may be identified with the space of single particle wave functions,
and which depends on the specific quantum field. At the level of the free electromagnetic field one can 
start at the Hilbert space of transversal single photon states acted on by the direct sum of the unitary 
zero mass helicity 1 and -1 representations respectively (in the language of the classical by now Wigner-Mackey-Gelfand-Bargmann classification of irreducible unitary representations of the Poincar\'e group). In more physical terms the representation
has been described e.g in \cite{bialynicki-2} together with its relation to the Riemann-Silbertstein vector wave 
function, and e.g. in \cite{amrein} in the form more closely related to the 
Wigner-Mackey scheme. It is true that the (free) quantum electromagnetic field has its own peculiarities
making some differences in comparison to massive and non gauge fields. The first peculiarity of a zero mass quantum
(free) field, even non gauge field (as we assume for a while in order to simplify situation), is that now the representation
of the Poincar\'e group to which we apply Segal's functor of second quantization although being unitary in ordinary sense, is specified within the 
Wigner-Mackey classification scheme by the orbit in the momentum space which is the light cone (without the apex),
contrary to the massive case, where the orbit is the smooth sheet of the two-sheeted hyperboloid. The apex being a singular point of the cone (in the sense of the ordinary differential structure of the cone as embedded into the $\mathbb{R}^4$-manifold) causes serious difficulties of infra-red character. This is because the quantum field is 
in fact an operator-valued distribution (as motivated by the famous Bohr-Rosenfeld  analysis \cite{bohr-rosenfeld} of the measurement of the quantum electromagnetic field) which needs a test function space. It is customary to use the standard Schwartz space of rapidly decreasing
functions as the universal test space even for zero mass fields, and this is not the correct test space for zero mass field if it is supposed to be constructed with the help of annihilation-creation generalized operators (at fixed momenta) which have the rigorous meaning of white noise Hida operators. Note in particular that it is not the case for mass less field in the sense of Wightman which uses ordinary Schwartz test space, but his definition of field is useless in realistic causal perturbative QFT. On the other hand the white noise construction of free fields is crucial in the causal perturbative approach of Bogoliubov-Epstein-Glaser.
Recall that the construction mentioned to above of a free quantum field achieved by the second 
quantization functor $\Gamma$ applied to a representation specified by a 
fixed orbit allows to construct creation and annihilation families of \emph{ordinary operators} in the Fock space. In order to construct the field as operator valued distribution (or generalized operator in the white noise sense of Berein-Hida) we have to proceed much further then in the construction given by Streater and Wightman in their well known 
monograph \cite{wig}, Ch. 3. In the construction of Wightman we consider the restrictions of Fourier transforms (i.e. functions in the momentum space) of the test functions to the orbit in question. The construction works for the field construced throgh white noise Hida operators if the restriction is a continuous map from the test function space in $\mathbb{R}^4$ to the test function space in 
$\mathbb{R}^3$ which is really the case in the massive case as the orbit is a smooth manifold in that case. 
Unfortunately it seems that it has escaped due attention of physicists that 1) the white noise construction of free fields is crucial for the causal perturbative QFT and 2) that  the correct test function space
in the momentum representation for the zero mass field (if constructed with the help of Hida annihilation-creation operators) should be equal to the closed subspace $\mathcal{S}_0$ of the Schwartz space 
$\mathcal{S}$ of those functions
which vanish at zero together with all their derivatives and the test function space $\mathcal{S}_{00}$ in the position representation is given by the inverse Fourier image of the space $\mathcal{S}_0$. This in turn causes additional 
difficulties concerned with exploring and correct use of the principles of locality character, because in particular the
space $\mathcal{S}_{00}$ does not contain any function of compact support (except the trivial zero function) which immediately follows from the generalized Paley-Wiener theorem. But the splitting of causal distributions works still well 
because the pairing functions of free fields are homogeneous, and the test space $\mathcal{S}_{00}$ is flexible enough to provide the basis for the splitting of causal and homogeneous distributions into the retarded and advanced parts.
$\mathcal{S}_{00}$ is also flexible enough to distinguish conic-type subsets and in particular for the explorance 
of the casality relations needed for the perturbative construction of the scattering matrix in the causal approach
of Bogoliubov-Epstein-Glaser.

The situation for the electromagnetic field is still more delicate as the field is accompanied by the gauge freedom and the ordinary unitarity is untenable
and has to be replaced with a weaker condition of preservation of the indefinite Krein-inner product
-- which is the second main peculiarity of the electromagnetic field, shared with the other zero mass 
gauge fields of the standard model.

Namely, although we may construct (remembering that we have to be careful with the choice of the test function space) the free quantum electric and magnetic fields by the mentioned 
application of the Segal second quantization functor to the direct sum of zero mas helicity 1 and -1 unitary representations
acting on the Riemann-Silberstein vector function (as described in \cite{bialynicki-2} or \cite{amrein}),
we encounter in this way a difficulty if we would like to restore the connection to the quantum vector potential
and its local transformation law within the the scheme. In principle we may reconstruct the quantum vector potential
in the momentum picture quite easily, but in connection to the non-local relationship of the potential to the electric and the magnetic fields in the position picture the local character of the vector potential is lost. 
We regard this a weakness when passing to interacting fields, specifically in passing to perturbative QED, and let us explain shortly why this is so. After half a century the causal method of St\"uckelberg and Bogoliubov turned up to be very valuable in avoiding the ultraviolet divergences in perturbative QFT, compare \cite{Epstein-Glaser}. Their method have been extended on QED and the other gauge fields, compare \cite{BlaSen}, \cite{DKS1}, \cite{DKS2}, \cite{DKS3}, \cite{DKS4}. A crucial
point of the method is the locality principle (local dependence of the interacting fields on the interaction Lagrangian,
\cite{DutFred}), and the second circumstance is that we need to have the quantum
vector potential -- recall that the minimal coupling is expressed immediately with the vector potential. Joining this 
prerequisites together we see that we need the vector potential with its local transformation law retained.
In particular we can achieve this within the Lorentz gauge with the four vector character of the transformation law of the quantum vector potential. However unitarity will have to be abandoned (recall the Gupta-Bleuler 
quantization \cite{Bleuler}).
Indeed, it is known that the four vector transformation law together with the zero mass character of the field cannot
be retained together with unitarity of the representation in the single particle space, compare e.g.
\cite{lop1}, \cite{lop2}. Because we prefer to stay within the micro-local perturbation scheme of QED and other gauge fields
of the standard model, avoiding ultraviolet divergences, we choose to abandon unitarity of the representation in the single photon states and replace Hilbert space and unitarity with Krein space and Krein-isometry property of the representation.
I.e. we  now have an ordinary Hilbert space together with two orthogonal projections $P_+$ and $P_-$
(of infinite dimensional ranges in our case) summing up to unity: $P_+ + P_- = I$, together with the 
fundamental symmetry $\mathfrak{J} = P_+ - P_-$
which in case of the Krein space of the free quantum electromagnetic filed is called the Gupta-Bleuler operator $\eta$.
Exactly as the ordinary Hilbert space structure and unitary representation admits the operation 
of direct sum and tensoring also the Krein space structure and Krein isometric representation preserving
the Krein inner product $(\cdot, \mathfrak{J} \cdot)$ (where $(\cdot, \cdot)$ is the ordinary 
Hilbert space inner product) admits direct summation and tensoring. In order to work effectively with such Krein-isometric representations we need to built a theory which plays the role analogous to the Mackey theory of induced representations.
We have constructed such a theory in \cite{wawrzycki-mackey} and in particular we have proved the main theorems,
namely we proved the Kronecker product theorem, subgroup theorem and the imprimitivity system theorem to hold for Krein-isometric representations induced by Krein-unitary\footnote{Krein-unitary operator is defined as a bouded operator preserving the Krein-inner-product possessing the bounded inverse which likewise preserves the Krein-inner-product. Krein isometry in general may be unbounded, but the representors of the Krein-isometric representations induced by Krein-unitary representations are closable on a common dense domain and are invertible on the common domain with the inversions being likewise closable and isometric for the Krein-inner-product, compare \cite{wawrzycki-mackey}.} representations
 (additional analytic assumptions are sufficiently weak to be effective
for physical applications). As an application we construct the free quantum electromagnetic field using the 
symmetrized tensoring and direct summation to a specific single photon indecomposable (but reducible)
Krein-isometric representation of the double covering of the Poincar\'e group, which we we call 
\emph{{\L}opusza\'nski representation}. We call the representation with {\L}opusza\'nski's name because he was the theoretician who suggested generalization of the Mackey theory in Krein spaces and strongly advocated the idea of construction of quantum electromagnetic field based on the representation theory in Krein spaces, 
compare e.g. \cite{lop1}, \cite{lop2}. In short we apply the second quantization functor $\Gamma$
to the \emph{{\L}opusza\'nski representation} in order to construct the free electromagnetic field.
In particular the operator $\eta = \Gamma(\mathfrak{J})$, where $\mathfrak{J}$ is the fundamental symmetry of the 
single photon representation (i.e. \emph{{\L}opusza\'nski representation}), is indeed equal to the Gupta-Bleuler operator. 
 
It is our aim of the paper to show that the construction of the transversal single photon space 
$\mathcal{H}_{\textrm{tr}}$ 
acted on by the  direct sum $[0,1] \oplus[0,-1]$ of unitary zero mass, helicty 1 and of helicity -1 representation as described in \cite{bialynicki-2} may be reconstructed as the closed subspace of physical states of the single photon 
Krein space with the representation $[0,1] \oplus[0,-1]$ induced by the action modulo unphysical states
of the single photon Krein-isometric representation. In this sense we extend the results on single photon wave
function, e.g. the uncertainty relations for the energy of single photon states obtained by prof. Bia{\l}ynicki
(compare \cite{bialynicki-2} and references therein) in showing their compatibility with the causal perturbative 
 QED, \cite{DutFred}, \cite{Bogoliubov_Shirkov}.

The paper is organized as follows. In Section 2 we define the single photon Krein-isometric representation, 
i.e. {\L}opusza\'nski
representation, as the representation induced by a Krein-unitary representation of the subgroup equal to the double covering of the euclidean plane, which preserves a zero vector (construction is based on the general theory presented in \cite{wawrzycki-mackey}) and concentrated on the light cone in momentum space. 
Although the representation preserves the Krein-inner product it is sufficiently singular,
being for example unbounded (each boost is represented by a closable unbounded operator), to require a good deal 
of care. For this reason we have presented all technical details separately in \cite{wawrzycki-mackey}.
In the same Section we construct a Krein-isometric representation which is Krein-equivalent (equivalence is given by a
Krein isometry preserving core dense domains of the representations which is invertible on the domains)
with the {\L}opusza\'nski representation
and has the property that after (inverse) Fourier transforming the single photon states in the momentum picture to the position picture we obtain local transformation law (in the specific case of the 
{\L}opusza\'nski representation it is just the four vector transformation law). In doing this we generalize the known construction of representations with local transformation law using our general theory \cite{wawrzycki-mackey},
which subsumes the construction for arbitrary spins and ordinary unitary representations as recapitulated e.g. 
in \cite{Haag}, I.3.3 pp. 31-32 as well as
in \cite{amrein}, and includes in particular Krein-isometric representations of (the double cover of) the Poincar\'e
group concentrated on single orbits in the momentum space. The general construction of the local single particle states using the results of \cite{wawrzycki-mackey} is shortly recapitulated in the Appendix for the reader's convenience.
In Section \ref{transversal} we present the construction of the closed physical subspace $\mathcal{H}_{\textrm{tr}}$ of transversal photons
together with the unitary representation acting on $\mathcal{H}_{\textrm{tr}}$ which is naturally generated
by the action, modulo unphysical states, of the single photon Krein-isometric representation, i. e. 
by the action of the {\L}opusza\'nski representation.

\section{{\L}opusza\'nski representation as the single photon representation}

Theory of induced representations presented in \cite{wawrzycki-mackey} is effective 
in treatment of a class of Krein-isometric representations of semi-direct locally compact 
products $G_1\circledS G_2$ of locally compact groups $G_1$ and $G_2$ with $G_1$ being abelian,
acting in a Krein space $(\mathcal{H}, \mathfrak{J})$. Krein space is nothing more but an ordinary Hilbert space
$\mathcal{H}$ endowed with the fundamental symmetry $\mathfrak{J}^2 = I$, $\mathfrak{J}^* = \mathfrak{J}$,
and Krein isometric representation preserves the Krein-inner-product $(\cdot, \mathfrak{J} \cdot)$, but
for detailed definition compare Sect. 2 of \cite{wawrzycki-mackey} as the peculiarities like unboundedness (with respect
to the ordinary Hilbert space product) cannot be excluded from the outset here in contrast to the ordinary unitary representations.
The class includes Krein-isometric representations of the double covering $T_4 \circledS SL(2, \mathbb{C})$
of the Poincar\'e group which we need in construction of the free electromagnetic field (and the other zero mass 
gauge fields of the standard model).  
We would like to avoid technicalities here but it is important to give some general feelings as to the additional 
analytic assumptions which allow us to work effectively with the special class of Krein-isometric representations
which are of interest in QFT. It should be stressed that the
general theory of Krein-isometric representations is substantially reacher and much more singular than the theory of ordinary unitary representations. In fact no general theory of such representations has been constructed even for
special classical groups which behave regularly (say of Type I). Neumark constructed a series of Krein-unitary representations of the Lorentz group \cite{Neumark-Pontriagin-Lorentz}
and proved some general theorems  on Krein-unitary representations\footnote{Krein-unitary but not unbounded Krein-isometric.} \cite{Neumark1}, \cite{Neumark2}, \cite{Neumark3}, \cite{Neumark4}, but only for the case where 
the fundamental symmetry 
$\mathfrak{J} = P_+ - P_-$ has $P_+$ (or $P_-$) with finite dimensional range\footnote{The case in which the Krein space degenerates to the so called Pontriagin space.} which unfortunately is 
useless in QFT. It is therefore important to understand the circumstances (supported by physics) which allows us to treat
the case in which both $P_+$ and $P_-$ have infinite rage. 

The first circumstance is that within the class of Krein-isometric representations which are of interest we may assume
that the representation restricted to the abelian normal subgroup $G_1$ (which we may identify with translations,
let us denote the restriction by $T$)
commutes with the fundamental symmetry $\mathfrak{J}$ (recall the Gupta-Bleuler formalism which at the naive but suggestive level gives momentum operator commuting with the Gupta-Bleuler operator $\eta$):
\begin{equation}\label{circumstance}
\mathfrak{J}T(a) = T(a)\mathfrak{J}, \,\,\, a \in G_1.
\end{equation}
 This circumstance is very
important and in particular means that the restriction $T(a)$, $a \in G_1$, of the Krein-isometric representation of 
$G_1\circledS G_2$ to the abelian normal factor $G_1$ (i.e. to the translations) is not only Krein-isometric but also preserves the ordinary Hilbert space inner-product of $\mathcal{H}$;
or in other words the Krein-isometric representation of $G_1\circledS G_2$, restricted to abelian normal factor $G_1$, 
gives an ordinary unitary
representation of $G_1$. This opens us to the full power of the theory of duality for locally compact abelian groups,
in particular we may apply the Neumark's theorem about the bi-unique correspondence between unitary representations
of locally compact abelian groups $G_1$ and spectral measures $E$ on their Pontriagin duals 
$\widehat{G_1}$ (or when we think
about the translation group $G_1 = T_4$ we may identify representations of $T_4$ with spectral measures in momentum space,
via the Fourier transform realizing the duality). This spectral measure $E$ gives the corresponding direct
integral decomposition (if one think of $G_1 = T_4$ being the normal factor of $T_4 \circledS SL(2,\mathbb{C})$) 
of the Hilbert space $\mathcal{H}$:  
\begin{equation}\label{circumstance1}
\mathcal{H} = \int \limits_{sp(P^0 , \ldots , P^3)} \mathcal{H}_{p} \, \ud \mu (p).
\end{equation}

The second circumstance is that we may restrict attention to such Krein-isometric representations 
of semi-direct products $G_1\circledS G_2$ for which the 
restriction $U(\alpha)$, $\alpha \in G_2$ to the second factor $G_2$ (we may identify $G_2$
with $SL(2,\mathbb{C})$) is locally bounded with respect to the above mentioned spectral measure $E$,
determined by the restriction $T(a)$, $a \in G_1$, of the representation of $G_1\circledS G_2$ to the abelian normal 
factor $G_1$, giving the decomposition  (\ref{circumstance1}). More precisely: let $\| \cdot \|$
be the ordinary Hilbert space $\mathcal{H}$ norm, then for every compact subset $\Delta$ of the 
dual $\widehat{G_1}$ and every $\alpha \in G_2$ there exists a positive constant 
$c_{\Delta, \alpha}$ (possibly depending on $\Delta$ and $\alpha$) such that
\begin{equation}\label{circumstance2}
\|U(\alpha) f \| < c_{\Delta, \alpha} \|f \|,
\end{equation}   
for all $f\in \mathcal{H}$ whose spectral support (in the spectral decomposition (\ref{circumstance1})
corresponding to $E$) is contained within the compact set $\Delta$. This means that all states with
bounded momentum whose ordinary Hilbert space norm have common bound are transformed by each
$U(\alpha)$ into states whose ordinary norm have common bound. This property is shared by the {\L}opusza\'nski
representation, but its justification is not so easily visible. 
In our opinion the only reasonable argument for its justification
lies in the proof that indeed the application of the Segal second quantization functor $\Gamma$
to the {\L}opusza\'nski representation gives the free quantum electromagnetic field, the full proof of which we publish as a separate paper as it is of much more mathematical character, and this is the argument to which we adhere 
in this paper.   

Under the circumstances (\ref{circumstance}) and (\ref{circumstance2})  Mackey's theory of induced 
representations may be generalized in treating
Krein-isometric representations of semi-direct products $G_1\circledS G_2$. 
In particular if we assume in addition that the semi-direct product is regular --
for example $T_4 \circledS SL(2,\mathbb{C})$ meets this requirement -- in the sense of Mackey (which means that 
every ergodic measure on $\widehat{G_1}$ with respect to the natural action of $G_2$ is concentrated on some
single orbit in $\widehat{G_1}$ with respect to this action), then we
may construct tensor product of such representations and compute effectively their direct integral  
decompositions into generalized induced representations concentrated on single orbits. 

In particular any
Krein-isometric representation of $T_4 \circledS SL(2,\mathbb{C})$ which meets the above stated requirements
(\ref{circumstance}) and (\ref{circumstance2})
may be decomposed into direct integral of representations such that for each of them the spectral measure $E$ is concentrated on a single orbit in $\widehat{T_4}$ -- which we identify via the Fourier transform with
the momentum space -- which is ergodic and invariant under the natural action of 
$SL(2, \mathbb{C})$ on  $\widehat{T_4}$. Let 
$\bar{p} \in \widehat{T_4}$ be any fixed point on the orbit $\mathscr{O}_{\bar{p}}$. Then, regarding its topology 
and Borel (or Baire) structure, $\mathscr{O}_{\bar{p}}$ is naturally isomorphic to the quotient group
$SL(2,\mathbb{C})/G_{\bar{p}}$, where $G_{\bar{p}}$ is the subgroup of $SL(2, \mathbb{C})$ of those elements
whose action on $\bar{p} \in \widehat{T_4}$ gives again $\bar{p}$, i.e. subgroup stationary for $\bar{p}$.
Let $\chi_{\bar{p}}$ be the character in $\widehat{T_4}$ determined by the momentum $\bar{p}$.
Form the subgroup $T_4 \cdot G_{\bar{p}}$ of $T_4 \circledS SL(2,\mathbb{C})$ consisting of all products 
$a \cdot \gamma$ with $a \in T_4$ and $\gamma \in G_{\bar{p}}$. For each Krein-unitary representation 
$L$ of the subgroup $G_{\bar{p}}$ the correspondence 
$a \cdot \gamma \mapsto \chi_{\bar{p}}(a) L(\gamma) = {}_{{}_{\chi_{\bar{p}}}} L(a \cdot \gamma)$,
defines a Krein-unitary representation of $T_4 \cdot G_{\bar{p}}$.  
It turns out that any Krein-isometric representation of $T_4 \circledS SL(2,\mathbb{C})$ preserving
(\ref{circumstance}) and (\ref{circumstance2}) which is concentrated on single orbit determined
by a closed subgroup $G_{\bar{p}}$ has the form (compare Appendix)
\begin{equation}\label{U,T,introduction}
\begin{split}
U(\alpha)\widetilde{\psi}(p) = Q(\gamma(\alpha,p),\bar{p}) \widetilde{\psi} (\Lambda(\alpha)p), \\
T(a) \widetilde{\psi}(p) = e^{i a \cdot p} \widetilde{\psi}(p), 
\end{split}
\end{equation} 
where $p \mapsto \widetilde{\psi}(p)$ are the decomposition functions of the elements $\widetilde{\psi} \in \mathcal{H}$
with respect to the decomposition (\ref{circumstance1}) determined by the spectral measure $E$
uniquely corresponding to the representation $T$, with the integral (\ref{circumstance1})
concentrated on the orbit $\mathscr{O}_{\bar{p}}$ and with the measure $\ud \mu (p)$ equal to the 
essentially unique measure $\ud \mu |_{{}_{\mathscr{O}_{\bar{p}}}} (p)$ on $\mathscr{O}_{\bar{p}}$
invariant with respect to the natural action of $SL(2,\mathbb{C})$; and where
$\gamma \mapsto Q(\gamma,\bar{p})$ is a Krein-unitary representation of the subgroup $G_{\bar{p}}$
and $p \mapsto \gamma(\alpha,p) \in G_{\bar{p}}$ a smooth function on the orbit with the values in 
$G_{\bar{p}}$ (i.e. $Q(\gamma,\bar{p})$ being a Krein unitary operator in the Krein space of the 
Krein-isometric representation of $G_{\bar{p}}$). Note that in general Krein isometric  
representation of $T_4 \circledS SL(2,\mathbb{C})$ concentrated on single orbit with the properties
(\ref{circumstance}) and (\ref{circumstance2}) has the property that its restriction $T$ to the abelian normal
factor has uniform (generally) infinite multiplicity (compare \cite{wawrzycki-mackey}).    

It turns out (Sect. 5 of \cite{wawrzycki-mackey}) that the Krein-isometric
representation (\ref{U,T,introduction}) of $T_4 \circledS SL(2,\mathbb{C})$ 
is Krein-unitary and unitary equivalent to the Krein-isometric 
representation $U^{{}_{{}_{\chi_{\bar{p}}}} L}$  of $T_4 \circledS SL(2,\mathbb{C})$ induced 
from the Krein-unitary representation ${}_{{}_{\chi_{\bar{p}}}} L$ of the subgroup
$T_4 \cdot G_{\bar{p}}$ in the sense defined in Sect. 2 
of \cite{wawrzycki-mackey} with $L(\cdot) = Q(\cdot,\bar{p})$. 

All these facts about the Krein-isometric representations of the indicated class
are very exceptional among Krein-isometric representations, in being closed with respect to the
operations of tensoring and of course direct summation, and especially their decomposability properties are very 
exceptional. Recall in particular that in general the existence of an (even closed) invariant subspace in the Krein space
does not guarantee decomposability of the Krein-isometric representation and neither existence of the Krein-self-adjoint
and bounded idempotent defining the invariant subspace is guaranteed.

In this paper we are interesting with a specific Krein-isometric representation
of $T_4 \circledS SL(2,\mathbb{C})$ induced by 
a finite dimensional Krein-unitary representation $L$ of the subgroup $G_{(1,0,0,1)}$ stationary
for the null vector $\bar{p} = (1,0,0,1)$. In other words the representation $T$ has finite uniform
multiplicity, so that the decomposition components (in the sense of von Neumann) may just be regarded
as ordinary functions with values in a finite dimensional Hilbert space $\mathcal{H}_{\bar{p}}$ which together
with a fundamental symmetry $\mathfrak{J}_{\bar{p}}$ composes the finite dimensional Krein space
of the Krein-unitary representation $L$ of the subgroup $G_{(1,0,0,1)}$. We may therefore identify the Hilbert space
$\mathcal{H}$ of the representation just with square integrable finite component vector functions with
respect to the invariant measure  
\[
\ud \mu |_{{}_{\mathscr{O}_{\bar{p}}}} (\vec{p}) = \frac{\ud^3 \vec{p}}{2  p^0 |_{{}_{\mathscr{O}_{\bar{p}}}}(\vec{p})}
= \frac{\ud^3 \vec{p}}{2 \sqrt{m^2 + \vec{p} \cdot \vec{p}}}.
\]
which in case of the null vector $\bar{p} = (1,0,0,1)$, i.e. with $\mathscr{O}_{\bar{p}}$ being just the null cone,
is equal
\[
\ud \mu |_{{}_{\mathscr{O}_{\bar{p}}}} (\vec{p})  
= \frac{\ud^3 \vec{p}}{2 |\vec{p}|}.
\] 

The disadvantageous property of the general Krein-isometric (and the same concerns ordinary unitary representation, 
when $\mathfrak{J} = I$) representation concentrated on a single orbit is that the multiplier in the formula
(\ref{U,T,introduction}) depends non-trivially on the momentum $p \in \mathscr{O}_{\bar{p}}$. This means that in the position
picture, i.e. after the application of the Fourier transform, we obtain non local transformation formula, so that the representation after the application of the second quantization functor $\Gamma$ gives a quantum field with
a non local transformation law. 

The task of constructing representation which in momentum space have a multiplier in the formula
(\ref{U,T,introduction}) which does not depend on $p$ and give after Fourier transforming a local transformation law
has been undertaken for ordinary unitary representations by several authors, compare \cite{Haag}, I.3.3
or \cite{amrein}. Using our previous results \cite{wawrzycki-mackey} we construct for a wide class
of Krein-isometric representations of $T_4 \circledS SL(2,\mathbb{C})$ fulfilling (\ref{circumstance})  
and (\ref{circumstance2}) a Krein-unitary and unitary operator $W$ which transforms
the initial Krein-isometric representation (\ref{U,T,introduction}) $(T, U)$ into Krein-unitary and 
unitary-equivalent representation $(WTW^{-1}, WUW^{-1})$ = $(T, WUW^{-1})$ of the form
\[
\begin{split}
WU(\alpha)W^{-1} \widetilde{\varphi} (p) = V(\alpha) \widetilde{\varphi} (\Lambda(\alpha)p), \\
WT(a)W^{-1} \widetilde{\varphi}(p) 
= e^{i a \cdot p}\widetilde{\varphi}(p),
\end{split}
\] 
for which the multiplier
in the formula (\ref{U,T,introduction}) is an operator (matrix) independent of $p$ and thus 
after the application of the Fourier transform we obtain local transformation formula. This construction
is in particular possible for the {\L}opusz\'nski representation and uses a Krein-unitary representation
$V$ of $SL(2,\mathbb{C})$ acting in the space of the representation $\gamma \mapsto Q(\gamma, \bar{p})$
in (\ref{U,T,introduction}) of the subgroup $G_{\bar{p}}$ and extends the representation 
$\gamma \mapsto Q(\gamma, \bar{p})$ to the whole $SL(2, \mathbb{C})$ group.

Construction  of this representation and its equivalent version giving local transformation in position picture
 may be treated as an example of the general
construction of local wave function presented in the Appendix based on \cite{wawrzycki-mackey}, which  
extends the results of \cite{amrein} and \cite{Haag}, I.3.3, to induced Krein-isometric representations in Krein spaces. 

Consider the orbit $\mathscr{O}_{(1,0,0,1)}$ of $\bar{p} = (1,0,0,1)$, i.e. positive energy surface of the cone (without
the apex $(0,0,0,0)$). The subgroup $G_{(1,0,0,1)} \subset SL(2, \mathbb{C})$ of matrices
\[
\gamma = (z, \phi) = \left( \begin{array}{cc} e^{i\phi/2} & e^{i \phi/2}z \\

                                     0 & e^{-i\phi/2}  \end{array}\right), \,\,\,
 0 \leq \phi < 4\pi , \,\,\, z \in \mathbb{C} 
\]
is stationary for $(1,0,0,1)$ and is isomorphic to the double covering group\footnote{Equal to the semi-direct
product $T_2 \circledS \widetilde{\mathbb{S}^1}$ of the two dimensional translation group $T_2$ and the double covering of
the circle group $\mathbb{S}^1$.} $\widetilde{E_2}$ of the Euclidean group $E_2$ of the Euclidean plane. 

As is well known there are no irreducible unitary representations of $G_{(1,0,0,1)}$ besides the infinite dimensional,
induced by the characters of the abelian normal subgroup $T_2$ of $G_{(1,0,0,1)}$ (numbered by a positive real number), 
and the one dimensional induced by the characters of the abelian subgroup $\widetilde{\mathbb{S}^1}$ and obtained by 
lifting to $G_{(1,0,0,1)}$ the one dimensional character representations of $G_{(1,0,0,1)}/T_2 \cong 
\widetilde{\mathbb{S}^1}$. And no standard combinations performed on them (direct summation, tensoring, conjugation)
can produce after a natural extension $V$ of the resulting representation to the whole $SL(2, \mathbb{C})$ the representation giving the ordinary transformation of a real four vector in Minkowski space (after the natural homomorphic map connecting  $SL(2, \mathbb{C})$ to the homogeneous Lorenz group). This is to be expected as the transformation
of the four vector under the Lorentz group is not unitary. The situation is different when passing to Krein-unitary representations of $G_{(1,0,0,1)}$.

Namely consider the following representation ${\L}$ of $G_{(1,0,0,1)}$ 
\[
{\L}_\gamma = S ( \gamma \otimes \overline{\gamma}) S^{-1}, \gamma \in G_{(1,0,0,1)}, 
\]
in $\mathbb{C}^4$, where 
\[
S =
\left( \begin{array}{cccc} \sqrt{2} & 0 & 0 & \sqrt{2} \\
                                0 & \sqrt{2} & \sqrt{2} & 0     \\
                                0 & i\sqrt{2} & -i \sqrt{2} & 0  \\
                                     \sqrt{2} & 0 & 0 & -\sqrt{2}  \end{array}\right)
\] 
is unitary in $\mathbb{C}^4$, and where $\overline{\gamma}$ means the ordinary complex conjugation:
if 
\[
\gamma
= \left( \begin{array}{cc} a & b \\ 
                                     c & d  \end{array}\right), \,\,\, \textrm{then} \,\,\,
\overline{\gamma}
= \left( \begin{array}{cc} \overline{a} & \overline{b} \\ 
                                     \overline{c} & \overline{d}  \end{array}\right).
\] 
If we introduce to $\mathbb{C}^4$ the ordinary inner product and the following fundamental symmetry operator
\begin{equation}\label{J-barp}
\mathfrak{J}_{\bar{p}} = \mathfrak{J}_{(1,0,0,1)} =
\left( \begin{array}{cccc} -1 & 0 & 0 & 0 \\
                                0 & 1 & 0 & 0     \\
                                0 & 0 & 1 & 0  \\
                                     0 & 0 & 0 & 1  \end{array}\right)
\end{equation}  
then the representation ${\L}$ of $ G_{(1,0,0,1)} = G_{\bar{p}}$ becomes Krein-unitary in the Krein space
$(\mathbb{C}^4, \mathfrak{J}_{\bar{p}})$:
\[
{\L} \, \mathfrak{J}_{\bar{p}} \, {\L}^* \, \mathfrak{J}_{\bar{p}} = \bold{1}_4, \,\,\, \textrm{and} \,\,\, 
\mathfrak{J}_{\bar{p}} \, {\L}^* \, \mathfrak{J}_{\bar{p}} \, {\L}  = \bold{1}_4, 
\]
where ${{\L}_\gamma}^*$ denotes the ordinary adjoint operator of ${\L}_\gamma$ with respect to the ordinary inner product
in $\mathbb{C}^4$.

The function $p \mapsto \beta(p) \in SL(2, \mathbb{C})$, fulfilling $\beta(p)^{-1} \, \widehat{\bar{p}} \, (\beta(p)^{-1})^{*} = \widehat{p}$
on the orbit $\mathscr{O}_{(1,0,0,1)}$, where $\widehat{p} = \widehat{(p^0, \ldots, p^3)} = p^0 \sigma_0 +
p^1 \sigma_1 + p^2 \sigma_2 + p^3 \sigma_3$ with Pauli matrices $\sigma_\mu$, may be chosen to be equal
\[
\beta(p) 
= \left( \begin{array}{cc} r^{{}^{-1/2}}\cos \frac{\theta}{2} e^{-i\frac{\vartheta}{2}} & 
-ir^{{}^{-1/2}}\sin \frac{\theta}{2} e^{i\frac{\vartheta}{2}} \\ 
 -ir^{{}^{1/2}}\sin \frac{\theta}{2} e^{-i\frac{\vartheta}{2}} & 
r^{{}^{1/2}}\cos \frac{\theta}{2} e^{i\frac{\vartheta}{2}}  \end{array}\right),
\] 
where
\begin{multline*}
p = 
\left( \begin{array}{c} p^0 \\
                               p^1     \\
                            p^2    \\
                               p^3       \end{array}\right) 
= \left( \begin{array}{c} r \\
                               r \sin \theta \sin \vartheta     \\
                            r \sin \theta \cos \vartheta     \\
                                r \cos \theta       \end{array}\right) \in \mathscr{O}_{(1,0,0,1)}, 
\,\,\, 0 \leq \theta < \pi, 
0 \leq \vartheta < 2 \pi, r > 0. 
\end{multline*}
Now we construct, like in the Appendix, the Krein-isometric representation of 
$T_4 \circledS SL(2, \mathbb{C})$ induced 
by the the Krein-unitary representation ${\L}$, putting there ${\L}_\gamma$ for $Q(\gamma, \bar{p})$
with $\bar{p} = (1,0,0,1)$. Let us denote the representation by $U^{{}_{(1,0,0,1)}{\L}}$
and call the \emph{{\L}opusza\'nski representation}. By Section 5 of \cite{wawrzycki-mackey}
(and the Appendix),
it is Krein-unitary equivalent to the Krein-isometric representation of $T^4 \circledS SL(2,\mathbb{C})$ 
induced\footnote{In the sense of definition placed in 
Section 2 of \cite{wawrzycki-mackey}, which is a generalization of the Mackey's induced representation.}
by the representation ${}_{(1,0,0,1)}{\L} = {}_{\chi_{\bar{p}}}{\L}$:
\[
a \cdot \gamma \mapsto \chi_{\bar{p}}(a) \, {\L}_\gamma, 
\]
of the subgroup $T_4 \cdot G_{\bar{p}} \subset T_4 \circledS SL(2,\mathbb{C})$. 

Now we define the following extension 
\[
V(\alpha) = S( \alpha \otimes \overline{\alpha}) S^{-1}, \alpha \in SL(2, \mathbb{C}),
\] 
of the representation ${\L}$, to the whole $SL(2, \mathbb{C})$ group, which is likewise 
Krein-unitary in $(\mathbb{C}^4, \mathfrak{J}_{\bar{p}})$:
\[
V(\alpha) \, \mathfrak{J}_{\bar{p}} \, {V(\alpha)}^* \, \mathfrak{J}_{\bar{p}} = \bold{1}_4, 
\,\,\, \textrm{and} \,\,\, 
\mathfrak{J}_{\bar{p}} \, {V(\alpha)}^* \, \mathfrak{J}_{\bar{p}} \, V(\alpha)  = \bold{1}_4, 
\,\,\, \alpha \in SL(2, \mathbb{C}).
\]
Moreover $\alpha \mapsto V(\alpha)$
gives a natural homomorphism of the $SL(2, \mathbb{C})$ onto the proper ortochronous Lorentz group
in the Minkowski vector space, i.e. each $V(\alpha)$, $\alpha \in SL(2, \mathbb{C})$, is a real Lorentz
transformation.
It is customary to write $V(\alpha)$ as the corresponding  Lorentz transformation $\Lambda(\alpha)$.
Because we have already occupied the notation $\Lambda(\alpha)$ for a natural antihomomorphism 
$\Lambda$, we have $V(\alpha) = \Lambda(\alpha^{-1})$ in our notation.  
 
With the extension $V$ at our disposal, we apply
to the elements $\widetilde{\psi}$  of the 
space  of the {\L}opusza\'nski representation $U^{{}_{(1,0,0,1)}{\L}}$  the Krein unitary and unitary
transformation $W: \widetilde{\psi} \mapsto \widetilde{\varphi}$ presented in the 
Appendix,
having the property that the Fourier transform (\ref{F(varphi)}) $\varphi$ has the local transformation
law. Namely the representation $WU^{{}_{(1,0,0,1)}{\L}}W^{-1}$ acts as follows
\begin{equation}\label{lop-rep-momentum}
\begin{split}
WU_{0,\alpha}^{{}_{(1,0,0,1)}{\L}}W^{-1} \widetilde{\varphi} (p) = 
U(\alpha) \widetilde{\varphi} (p) = V(\alpha) \widetilde{\varphi} (\Lambda(\alpha)p), \\
WU_{a,1}^{{}_{(1,0,0,1)}{\L}}W^{-1}  \widetilde{\varphi} (p) = T(a) \widetilde{\varphi}(p) 
= e^{i a \cdot p}\widetilde{\varphi}(p). 
\end{split}
\end{equation}
Therefore the Fourier transform (\ref{F(varphi)}) $\varphi$ of $\widetilde{\varphi} = W \widetilde{\psi}$ has the 
the following local transformation law
\[
\begin{split}
U(\alpha)\varphi(x) = V(\alpha) \varphi(x \Lambda(\alpha^{-1}))
= \Lambda(\alpha^{-1}) \varphi(x \Lambda(\alpha^{-1})), \,\,\,
T(a)\varphi(x) = \varphi(x - a).
\end{split}
\]
of a four vector field on the Minkowski manifold. Because by construction $\widetilde{\varphi}$ are concentrated
on the orbit $\mathscr{O}_{(1,0,0,1)}$, it follows that the elements $\varphi \in \mathcal{H}''$
are the positive energy (distributional) solutions of the ordinary wave equation with zero mass
\[
\partial^\mu \partial_\mu \varphi = 0.
\] 

The explicit form of the Krein space structure $(\mathcal{H}', \mathfrak{J}')$ of the representation space of the
representation $WU^{{}_{(1,0,0,1)}{\L}}W^{-1}$ can be obtained by substitution of the explicit formulas for 
the function $p \mapsto \beta(p)$ and the extension $V$ into the formulas written in the Appendix.
 
In particular the inner product of $\widetilde{\varphi} = W \widetilde{\psi}$ and $\widetilde{\varphi'} = 
W \widetilde{\psi'}$ is equal
\begin{multline}\label{inn-Lop-1-space}
(\widetilde{\varphi}, \widetilde{\varphi}') 
= \int \limits_{sp(P^0 , \ldots , P^3) \cong \mathscr{O}_{\bar{p}}} 
\Big( \widetilde{\varphi}(p), \widetilde{\varphi}'(p) \Big)_{p} 
\, \ud \mu |_{{}_{\mathscr{O}_{\bar{p}}}} (p) \\
= \int \limits_{sp(P^0 , \ldots , P^3) \cong \mathscr{O}_{\bar{p}}} 
\Big(  \widetilde{\varphi}(p), 
V(\beta(p))^* V(\beta(p)) \widetilde{\varphi}'(p)  \Big)_{\mathcal{H}_{\bar{p}}}
\, \ud \mu |_{{}_{\mathscr{O}_{\bar{p}}}} (p) \\
= \int \limits_{\mathscr{O}_{\bar{p}}} 
\Big(  \widetilde{\varphi}(p), 
B(p) \widetilde{\varphi}'(p)  \Big)_{\mathbb{C}^4}
\, \ud \mu |_{{}_{\mathscr{O}_{\bar{p}}}} (p), \\ 
= \int \limits_{\mathbb{R}^3} 
\Big(  \widetilde{\varphi}(\vec{p}, p^0(\vec{p})), 
(B\widetilde{\varphi}')(\vec{p}, p^0(\vec{p}))  \Big)_{\mathbb{C}^4}
\, \ud^3 p = (\widetilde{\varphi}, B \widetilde{\varphi}')_{{}_{\oplus L^2(\mathbb{R}^3)}} , \\   
\ud \mu |_{{}_{\mathscr{O}_{\bar{p}}}} (\vec{p}) = \frac{\ud^3 p}{2p^0(\vec{p})}, \, p^0(\vec{p}) = (\vec{p}\cdot \vec{p})^{{}^{1/2}},  
\end{multline}
where we have introduced the matrix 
\[
B(p) = V(\beta(p))^* V(\beta(p))
\]
depending on $p \in \mathscr{O}_{\bar{p}}$,  strictly positive (invertible) on $\mathscr{O}_{\bar{p}}$
and the operator $B$ of point-wise multiplication by the matrix
\begin{equation}\label{operatorB}
\frac{1}{2p^0(\vec{p})} B\big(\vec{p}, p^0(\vec{p})\big),
\end{equation}
on the Hilbert space $\oplus L^2(\mathbb{R}^3) = L^2(\mathbb{R}^3; \mathbb{C}^4)$ with respect to the ordinary invariant Lebesgue measure
$\ud^3 \boldsymbol{\p}$ on $\mathbb{R}^3$ (the direct sum $\oplus$ is over the four components of the 
function $\widetilde{\varphi}$), 
in order to simplify notation of the formulas which are to follow in the remaining part of this paper.

The fundamental symmetry operator $\mathfrak{J'}$ is given by the point-wise multiplication by the following 
operator
\[
\mathfrak{J'}_{p} = V(\beta(p))^{-1} \mathfrak{J}_{\bar{p}} V(\beta(p)).
\] 
Because  for each $p \in \mathscr{O}_{\bar{p}}$ the matrix operator  $V(\beta(p))$ (and the same of course 
holds for $V(\beta(p))^*$)   is by construction Krein-unitary 
in the Krein space $(\mathbb{C}^4, \mathfrak{J}_{\bar{p}})
 = (\mathcal{H}_{\bar{p}}, \mathfrak{J}_{\bar{p}})$ of the representation ${\L}$, 
then the Krein product in $(\mathcal{H}', \mathfrak{J'})$ is given by the following formula
\begin{multline}\label{Kr-inn-Lop-1-space}
(\widetilde{\varphi}, \mathfrak{J'} \widetilde{\varphi}') 
= \int \limits_{sp(P^0 , \ldots , P^3) \cong \mathscr{O}_{\bar{p}}} 
\Big( \widetilde{\varphi}(p), V(\beta(p))^* V(\beta(p)) \mathfrak{J'}_p \widetilde{\varphi}'(p)  
 \Big)_{\mathcal{H}_{\bar{p}}} 
\, \ud \mu |_{{}_{\mathscr{O}_{\bar{p}}}} (p) \\
= \int \limits_{\mathscr{O}_{\bar{p}}} 
\Big( \widetilde{\varphi}(p), V(\beta(p))^* V(\beta(p)) V(\beta(p))^{-1} \mathfrak{J}_{\bar{p}} V(\beta(p)) \widetilde{\varphi}'(p) \Big)_{\mathbb{C}^4} 
\, \ud \mu |_{{}_{\mathscr{O}_{\bar{p}}}} (p) \\
= \int \limits_{\mathscr{O}_{\bar{p}}} 
\Big( \widetilde{\varphi}(p), \mathfrak{J}_{\bar{p}} \widetilde{\varphi}'(p) \Big)_{\mathbb{C}^4} 
\, \ud \mu |_{{}_{\mathscr{O}_{\bar{p}}}} (p), \\
\end{multline}
because $V(\beta(p))^* \mathfrak{J}_{\bar{p}} V(\beta(p)) = \mathfrak{J}_{\bar{p}}$.

Introducing the coordinates $\vec{p}$ on $\mathscr{O}_{\bar{p}}$ and regarding any function
$p \mapsto \widetilde{\varphi}(p)$ on $\mathscr{O}_{\bar{p}}$ as a function 
$\vec{p} \mapsto \widetilde{\varphi}(\vec{p}) = \widetilde{\varphi}\big(\vec{p}, p^0(\vec{p})\big)$ with $p^0(\vec{p})$
as in (\ref{inn-Lop-1-space}), the last formula (\ref{Kr-inn-Lop-1-space}) may be written as
\begin{multline}\label{Kr-inn-Lop-1-space'}
(\widetilde{\varphi}, \mathfrak{J'} \widetilde{\varphi}') 
= (\widetilde{\varphi}, B \mathfrak{J}' \widetilde{\varphi}')_{{}_{\oplus L^2(\mathbb{R}^3)}}
= (\sqrt{B}\widetilde{\varphi}, \sqrt{B} \mathfrak{J}' \widetilde{\varphi}')_{{}_{\oplus L^2(\mathbb{R}^3)}} 
= (\widetilde{\varphi}, \mathfrak{J}_{\bar{p}} \widetilde{\varphi}')_{{}_{\oplus L^2(\mathbb{R}^3, 
\ud \mu |_{{}_{\mathscr{O}_{\bar{p}}}})}} ,
\end{multline}  
where the last inner product 
\[
(\cdot, \cdot)_{{}_{\oplus L^2(\mathbb{R}^3, \ud \mu |_{{}_{\mathscr{O}_{\bar{p}}}})}}
\]
is with respect to the measure
\[
\ud \mu = \frac{\ud^3 p}{2p^0(\vec{p})}, \,\, p^0(\vec{p}) = (\vec{p}\cdot \vec{p})^{{}^{1/2}},
\]
on $\mathbb{R}^3$, and where $B$ is the positive self-adjoint operator on $\oplus L^2(\mathbb{R}^3)$ introduced above and 
$\sqrt{B}$ is its square root equal to the operator of point-wise multiplication by the matrix
\[
\frac{1}{\sqrt{2p^0(\vec{p})}} \sqrt{B\big(\vec{p}, p^0(\vec{p})\big)},
\]

with $\sqrt{B\big(\vec{p}, p^0(\vec{p})\big)}$  being the square root of the positive matrix
$B\big(\vec{p}, p^0(\vec{p})\big)$.

Krein-isometric and Krein-unitary representations in a Krein space $(\mathcal{H}, \mathfrak{J})$ allows the specific 
kind of conjugation, which is trivial for ordinary unitary representations when $\mathfrak{J} = \bold{1}_\mathcal{H}$. Namely for every representation $U$ of this kind in the Krein space $(\mathcal{H}, \mathfrak{J})$,
the ordinary Hilbert space adjoint
operation $^*$ and passing to the inverse, i.e. $U^{*-1} = \mathfrak{J}U\mathfrak{J}$, is well defined, which is nontrivial 
for Krein-isometric representation, compare \cite{wawrzycki-mackey}, Sect. 2. Moreover $U^{*-1} = \mathfrak{J}U\mathfrak{J}$ defines another Krein-isometric (resp. Krein unitary)
representation with respect to the same Krein structure, compare \cite{wawrzycki-mackey}, Sect. 2, which is unitary and Krein-unitary equivalent to the initial representation $U$, with the equivalence given by the fundamental symmetry 
$\mathfrak{J}$
itself, and $\mathfrak{J}$ is by construction unitary and Krein-unitary. 

In particular together with the Krein-isometric representation $WU^{{}_{(1,0,0,1)}{\L}}W^{-1}$ in
the Krein space $(\mathcal{H}', \mathfrak{J}')$ just constructed, there acts in the same Krein space
$(\mathcal{H}', \mathfrak{J}')$ the naturally conjugate Krein isometric representation
\[
\big[WU^{{}_{(1,0,0,1)}{\L}}W^{-1}\big]^{*-1} = \mathfrak{J'} WU^{{}_{(1,0,0,1)}{\L}}W^{-1} \mathfrak{J'}
\]
unitary and Krein-unitary equivalent to $WU^{{}_{(1,0,0,1)}{\L}}W^{-1}$, with the equivalence
given by the fundamental symmetry $ \mathfrak{J'}$ itself. Because we have explicitly computed
$\mathfrak{J'}$ and $WU^{{}_{(1,0,0,1)}{\L}}W^{-1}$ we also know the explicit formula for the action of 
$\big[WU^{{}_{(1,0,0,1)}{\L}}W^{-1}\big]^{*-1}$. Namely we have 
\begin{multline*}
\big[WU^{{}_{(1,0,0,1)}{\L}}W^{-1}\big]^{*-1} \widetilde{\varphi}(p) =
\big( \mathfrak{J'} WU^{{}_{(1,0,0,1)}{\L}}W^{-1} \mathfrak{J'} \big) \widetilde{\varphi}(p) \\ 
= V(\beta(p))^{-1} V(\beta(p))^{*-1} V(\alpha)^{*-1} V(\beta(\Lambda(\alpha)p))^{*} 
V(\beta(\Lambda(\alpha)p)) \widetilde{\varphi}(\Lambda(\alpha)p).
\end{multline*}

Before passing to quantization, we give here several formulas which will be useful in further computations.
First let us note the simple formula for the Krein inner product in the Krein space 
$(\mathcal{H''}, \mathfrak{J''})$ of all  Fourier transforms $\varphi$, given by (\ref{F(varphi)}), of the
elements $\widetilde{\varphi}$ of the Krein space $(\mathcal{H'}, \mathfrak{J'})$ of the representation
$WU^{{}_{(1,0,0,1)}{\L}}W^{-1}$. Namely easy computation gives
\begin{multline}\label{krein-prod-1-photon}
(\varphi, \mathfrak{J''}\varphi) = 
i \int \limits_{t = const.} \Big\{
 \overline{\varphi(x)}\partial_t\big(\mathfrak{J}_{\bar{p}} \varphi'\big)(x) 
- \overline{\partial_t \varphi(x)} \mathfrak{J}_{\bar{p}} \varphi'(x) \Big\} \, \ud^3 x \\
= - i g^{\mu \nu} \int \limits_{t = const.} \Big\{
 \overline{\varphi_\mu (x)}\partial_t \varphi'_\nu\big(x) 
- \overline{\partial_t \varphi_\mu(x)} \varphi_\nu (x) \Big\} \, \ud^3 x
\end{multline}

Next we give explicit formulas for $V(\beta(p))^{-1}$, $B(p) = V(\beta(p))^* V(\beta(p))$ and 
$\sqrt{B(p)}$, $p \in \mathscr{O}_{(1,0,0,1)}$, and give their useful properties.
\begin{multline*}
V(\beta(p))^{-1} \\
= \left( \begin{array}{cccc} 
\frac{r^{-1} + r}{2} & 0 & 0 & -\frac{r^{-1} - r}{2} \\
-\frac{r^{-1} - r}{2} \frac{p^1}{r} & \frac{p^2}{\sqrt{(p^1)^2 + (p^2)^2}} &  \frac{p^1}{\sqrt{(p^1)^2 + (p^2)^2}} \frac{p^3}{r} & \frac{r^{-1} + r}{2} \frac{p^1}{r}  \\
-\frac{r^{-1} - r}{2} \frac{p^2}{r} & - \frac{p^1}{\sqrt{(p^1)^2 + (p^2)^2}} & \frac{p^2}{\sqrt{(p^1)^2 + (p^2)^2}}  \frac{p^3}{r} &  \frac{r^{-1} + r}{2} \frac{p^2}{r}  \\
-\frac{r^{-1} - r}{2} \frac{p^3}{r} & 0 & -\frac{\sqrt{(p^1)^2 + (p^2)^2}}{r} & \frac{r^{-1} + r}{2} \frac{p^3}{r} \end{array}\right) \\
= \left( \begin{array}{cccc} 
\frac{r^{-1} + r}{2} & 0 & 0 & -\frac{r^{-1} - r}{2} \\
-\frac{r^{-1} - r}{2} \sin \theta \sin \vartheta & \cos \vartheta & \cos \theta \sin \vartheta & \frac{r^{-1} + r}{2} \sin \theta \sin \varphi \\
-\frac{r^{-1} - r}{2} \sin \theta \cos \varphi & -\sin \vartheta & \cos \theta \cos \vartheta & \frac{r^{-1} + r}{2} \sin \theta \cos \vartheta  \\
-\frac{r^{-1} - r}{2} \cos \theta & 0 & - \sin \theta & \frac{r^{-1} + r}{2} \cos \theta \end{array}\right). 
\end{multline*}
\begin{multline}\label{Bmatrix}
B(p) = V(\beta(p))^* V(\beta(p)) = \\
\left( \begin{array}{cccc} 
\frac{r^{-2} + r^2}{2} & \frac{r^{-2} - r^2}{2r}p^1 & \frac{r^{-2} - r^2}{2r}p^2 & \frac{r^{-2} - r^2}{2r}p^3 \\
\frac{r^{-2} - r^2}{2r} p^1&\frac{r^{-2} + r^2 -2}{2r^2}p^1 p^1 +1 & \frac{r^{-2} + r^2 -2}{2r^2}p^1 p^2 & \frac{r^{-2} + r^2 -2}{2r^2} p^1 p^3  \\
\frac{r^{-2} - r^2}{2r}p^2 & \frac{r^{-2} + r^2 -2}{2r^2}p^2 p^1 &\frac{r^{-2} + r^2 -2}{2r^2}p^2 p^2 +1 & \frac{r^{-2} + r^2 -2}{2r^2} p^2 p^3  \\
\frac{r^{-2} - r^2}{2r}p^3 & \frac{r^{-2} + r^2 -2}{2r^2}p^3 p^1 & \frac{r^{-2} + r^2 -2}{2r^2}p^3 p^2 &\frac{r^{-2} + r^2 -2}{2r^2}p^3 p^3 +1  \end{array}\right) 
\end{multline}
\begingroup\makeatletter\def\f@size{5}\check@mathfonts
\def\maketag@@@#1{\hbox{\m@th\large\normalfont#1}}%
\begin{multline*}
= 
\left( 
\begin{array}{ccc} 
\frac{r^{-2} + r^2}{2} & \frac{r^{-2} - r^2}{2}\sin \theta \sin \vartheta &  \\
\frac{r^{-2} - r^2}{2}\sin \theta \sin \vartheta & \frac{r^{-2} + r^2}{2}\sin^2 \theta \sin^2 \vartheta +\cos^2 \theta \sin^2 \vartheta +\cos^2 \vartheta &  \ldots \\
\frac{r^{-2} - r^2}{2}\sin \theta \cos \vartheta & \frac{r^{-2} + r^2}{2}\sin^2 \theta \cos \vartheta \sin \vartheta +\cos^2 \theta \sin \vartheta \cos \vartheta - \sin \vartheta \cos \vartheta &   \\
\frac{r^{-2} - r^2}{2} \cos \theta & \frac{r^{-2} + r^2}{2} \sin \theta \cos \theta \sin \vartheta
- \sin \theta \cos \theta \sin \vartheta &   \end{array} \right. \\
\end{multline*}
\begin{multline*}
\,\,\,\,\,\,\,\,\,\,\,\,\,\,\,\,\,\,\,\,\,\,\,\,\,\,\,\,\,\,\,\,\,\,\,\,\,\,\,\,\,\,\,\,\,\,\,\,\,\,\,\,\,\,\,\,\,
\left.
\begin{array}{ccc} 
 & \frac{r^{-2} - r^2}{2}\sin \theta \cos \vartheta & \frac{r^{-2} - r^2}{2} \cos \theta \\
 \ldots & \frac{r^{-2} + r^2}{2}\sin^2 \theta \cos \vartheta \sin \vartheta +\cos^2 \theta \sin \vartheta \cos \vartheta - \sin \vartheta \cos \vartheta & \frac{r^{-2} + r^2}{2} \sin \theta \cos \theta \sin \vartheta
- \sin \theta \cos \theta \sin \vartheta \\
 & \frac{r^{-2} + r^2}{2}\sin^2 \theta \cos^2 \vartheta +\cos^2 \theta \cos^2 \vartheta +\sin^2 \vartheta & \frac{r^{-2} + r^2}{2} \sin \theta \cos \theta \cos \vartheta - \sin \theta \cos \theta \cos \vartheta  \\
 & \frac{r^{-2} + r^2}{2} \sin \theta \cos \theta \cos \vartheta - \sin \theta \cos \theta \cos \vartheta & \frac{r^{-2} + r^2}{2}p^3 \cos^2 \theta + \sin^2 \theta  \end{array} \right). 
\end{multline*}\endgroup   

The orthonormal (with respect to the ordinary inner product in $\mathbb{C}^4$) system $\{w_\lambda(p) \}$ of eigenvectors   of the operator matrix $B(p) =  V(\beta(p))^* V(\beta(p))$ in $\mathbb{C}^4$, corresponding to the eigenvalues
$\lambda(p) \in \{1,1, r^{-2}, r^2 \}$  has the form
\begin{multline*}
{w_{{}_1}}^+(p) = \left( \begin{array}{c} 0 \\
                               \frac{p^2}{\sqrt{(p^1)^2 + (p^2)^2}}     \\
                             \frac{-p^1}{\sqrt{(p^1)^2 + (p^2)^2}}   \\
                               0       \end{array}\right), 
{w_{{}_1}}^- (p)= \left( \begin{array}{c} 0 \\
                               \frac{p^1 p^3}{\sqrt{(p^1)^2 + (p^2)^2}r}     \\
                             \frac{p^2 p^3}{\sqrt{(p^1)^2 + (p^2)^2}r}   \\
                              - \frac{\sqrt{(p^1)^2 + (p^2)^2}}{r}       \end{array}\right), \\ 
w_{{}_{r^{-2}}}(p) = \left( \begin{array}{c} \frac{1}{\sqrt{2}} \\
                              \frac{1}{\sqrt{2}}\frac{p^1}{r}     \\
                           \frac{1}{\sqrt{2}}\frac{p^2}{r}    \\
                             \frac{1}{\sqrt{2}}\frac{p^3}{r}       \end{array}\right),
w_{{}_{r^2}}(p) = \left( \begin{array}{c} \frac{1}{\sqrt{2}} \\
                              -\frac{1}{\sqrt{2}}\frac{p^1}{r}     \\
                           -\frac{1}{\sqrt{2}}\frac{p^2}{r}    \\
                             -\frac{1}{\sqrt{2}}\frac{p^3}{r}       \end{array}\right) 
\end{multline*} 
There are two transversal eigenvectors ${w_{{}_1}}^+(p), {w_{{}_1}}^-(p)$ to the constant eigenvalue 1,
both of pure space direction and both orthogonal to the space part $(0,\vec{p})$ of the momentum direction of the 
corresponding momentum $p = (p^0, \vec{p}) \in \mathscr{O}_{(1,0,0,1)}$. The eigenvector
$w_{{}_{r^{-2}}}(p)$ corresponding to the eigenvalue $r^{-2} = (p^0)^{-2} = (\vec{p} \cdot \vec{p})^{-1}$, 
has the same direction as the 
corresponding momentum $p = (p^0, \vec{p}) \in \mathscr{O}_{(1,0,0,1)}$, and 
$w_{{}_{r^2}}(p)$ has the same direction as $(p^0, -\vec{p})$, where 
$p =(p^0, \vec{p}) \in \mathscr{O}_{(1,0,0,1)}$
is the corresponding momentum. Note that the linear combinations $w_{{}_{r^{-2}}}(p) + w_{{}_{r^2}}(p)$
and $w_{{}_{r^{-2}}}(p) - w_{{}_{r^2}}(p)$ give respectively the purely time-like vector
of direction the same as $(p^0, 0)$ and a purely longitudinal
vector of direction the same as $(0, \vec{p})$, where $p = (p^0, \vec{p}) \in \mathscr{O}_{(1,0,0,1)}$ is the corresponding
momentum vector. 

The square root of $B(p) =  V(\beta(p))^* V(\beta(p))$  is equal
\begin{multline}\label{sqrtB}
\sqrt{V(\beta(p))^* V(\beta(p))} = \sqrt{B(p)} \\
= \left( \begin{array}{cccc} 
\frac{r^{-1} + r}{2} & \frac{r^{-1} - r}{2} \frac{p^1}{r} & \frac{r^{-1} - r}{2} \frac{p^2}{r} & \frac{r^{-1} - r}{2} \frac{p^3}{r} \\
\frac{r^{-1} - r}{2} \frac{p^1}{r} & \frac{r^{-1} + r -2}{2} \frac{p^1}{r} \frac{p^1}{r} +1 &  \frac{r^{-1} + r -2}{2} \frac{p^1}{r} \frac{p^2}{r} & \frac{r^{-1} + r -2}{2} \frac{p^1}{r} \frac{p^3}{r} \\
\frac{r^{-1} - r}{2} \frac{p^2}{r} & \frac{r^{-1} + r -2}{2} \frac{p^2}{r} \frac{p^1}{r} & \frac{r^{-1} + r -2}{2}  \frac{p^2}{r} \frac{p^2}{r} +1 & \frac{r^{-1} + r -2}{2} \frac{p^2}{r} \frac{p^3}{r} \\
\frac{r^{-1} - r}{2} \frac{p^3}{r} & \frac{r^{-1} + r -2}{2} \frac{p^3}{r} \frac{p^1}{r} & \frac{r^{-1} + r -2}{2} \frac{p^3}{r}\frac{p^2}{r} & \frac{r^{-1} + r -2}{2} \frac{p^3}{r} \frac{p^3}{r} + 1\end{array}\right) 
\end{multline}
\begingroup\makeatletter\def\f@size{5}\check@mathfonts
\def\maketag@@@#1{\hbox{\m@th\large\normalfont#1}}%
\begin{multline*}
= \left( \begin{array}{cccc} 
\frac{r^{-1} + r}{2} & \frac{r^{-1} - r}{2} \sin \theta \sin \vartheta & \frac{r^{-1} - r}{2} \sin \theta \cos \vartheta & \frac{r^{-1} - r}{2} \cos \theta \\
\frac{r^{-1} - r}{2} \sin \theta \sin \vartheta & \frac{r^{-1} + r -2}{2} \sin^2 \theta \sin^2 \vartheta  +1 &  \frac{r^{-1} + r -2}{2} \sin^2 \theta \sin \vartheta \cos \vartheta & \frac{r^{-1} + r -2}{2} \sin \theta \cos \theta \sin \vartheta \\
\frac{r^{-1} - r}{2} \sin \theta \cos \vartheta & \frac{r^{-1} + r -2}{2} \sin^2 \theta \cos \vartheta \sin \vartheta & \frac{r^{-1} + r -2}{2}  \sin^2 \theta \cos^2 \vartheta +1 &  \frac{r^{-1} + r -2}{2} \sin \theta \cos \theta \cos \vartheta \\
\frac{r^{-1} - r}{2} \cos \theta & \frac{r^{-1} + r -2}{2} \sin \theta \cos \theta \sin \vartheta & \frac{r^{-1} + r -2}{2} \sin \theta \cos \vartheta \cos \vartheta & \frac{r^{-1} + r -2}{2} \cos^2 \theta +1 \end{array}\right). 
\end{multline*}\endgroup 

By construction $V(\beta(p))$, $V(\beta(p))^* = V(\beta(p))^{T}$ and their inverses are at every 
$p \in \mathscr{O}_{(1,0,0,1)}$ Krein unitary, as matrix operators in $(\mathcal{H}_{\bar{p}}, \mathfrak{J}_{\bar{p}})
= (\mathbb{C}^4, \mathfrak{J}_{\bar{p}})$, i.e they are real Lorentz transformations. It is less trivial, but may be checked directly that for every $p \in \mathscr{O}_{(1,0,0,1)}$ the operator $\sqrt{B(p)}$ is also Krein unitary
in $(\mathbb{C}^4, \mathfrak{J}_{\bar{p}})$. Thus we have the formulas
\begin{multline}\label{BJBJ=1}
V(\beta(p)) \, \mathfrak{J}_{\bar{p}} \, {V(\beta(p))}^* \, \mathfrak{J}_{\bar{p}} = \bold{1}_4, 
\,\,\, \textrm{and} \,\,\, 
\mathfrak{J}_{\bar{p}} \, {V(\beta(p))}^* \, \mathfrak{J}_{\bar{p}} \, V(\beta(p))  = \bold{1}_4,  \,\,\,\textrm{and} \\
\sqrt{B(p)} \, \mathfrak{J}_{\bar{p}} \, \sqrt{B(p)} \, \mathfrak{J}_{\bar{p}} = \bold{1}_4,
\,\,\, p \in \mathscr{O}_{(1,0,0,1)}.
\end{multline}

Although the properties are simple consequences of definitions (possibly with the exception of the last one)
they will be of use in further computations.

\begin{center}
\small DEFINITION OF THE KREIN-HILBERT SPACE $(\mathcal{H'}, \mathfrak{J'})$ WHICH IS THEN SUBJECT TO THE SECOND 
QUANTIZATION FUNCTOR $\Gamma$
\end{center}

Now to the Hilbert space $\mathcal{H'}$, or more precisely to the Krein space $(\mathcal{H'}, \mathfrak{J'})$
of the representation $WU^{{}_{(1,0,0,1)}{\L}}W^{-1}$ and \emph{eo ipso} of the representation 
\[
\big[WU^{{}_{(1,0,0,1)}{\L}}W^{-1}\big]^{*-1},
\] 
we apply the Segal's bosonic second quantization
functor $\Gamma$. The Krein space $(\mathcal{H'}, \mathfrak{J'}) = \big( W \mathcal{H}, W\mathfrak{J}W^{-1} \big)$
of the elements $\widetilde{\varphi} = W\widetilde{\psi}$ of the representation 
\[
WU^{{}_{(1,0,0,1)}{\L}}W^{-1}
\]
may be identified, via the Fourier transform (\ref{F(varphi)}) with the Hilbert space $\mathcal{H''}$, or more precisely
with the Krein space $(\mathcal{H''}, \mathfrak{J''})$ of positive energy solutions $\varphi$
of the wave equation
\begin{equation}\label{nonloc-evol-wave}
g^{\mu \nu}\partial_\mu \partial_\nu \varphi = 0,
\end{equation} 
as a consequence of the fact that $\widetilde{\varphi} \in \mathcal{H'}$ are concentrated on the cone $\mathscr{O}_{(1,0,0,1)} = \mathscr{O}_{(1,0,0,1)}$. Although it is well known that the equation (\ref{nonloc-evol-wave}) only 
apparently gives a local law for dynamics in terms of a local equation. Indeed because only the positive energy 
solutions\footnote{In the construction of the positive energy field via the second quantization functor applied 
to to the space $(\mathcal{H'}, \mathfrak{J'})$. In the construction of the negative energy field the roles of positive and negative energy is interchanged.} are admitted the quantities $\varphi$ and $\partial_t\varphi$ are not independent
on a fixed time surface.
The differentiation $\partial_t$ in momentum space is equal to the operator of multiplication by $-i \sqrt{\vec{p}\cdot \vec{p}}$, which in position picture at fixed time corresponds to a convolution with the non-local integral kernel\footnote{Already the definition of the kernel necessitates a special care, and may be defined in the distributional sense }
\[
K(\vec{x} - \vec{x'}) = -i (2 \pi)^{-3/2} \int \limits_{\mathbb{R}^3} \,
\sqrt{\vec{p}\cdot \vec{p}} \,\, e^{i \vec{p} \cdot (\vec{x} - \vec{x'})} \ud^3 p,
\]
exactly as for the spin-less massive particles (compare e. g. \cite{Haag}, I. 3.3.).

Unfortunately the ordinary Hilbert space inner product $(\varphi, \varphi')  = (\widetilde{\varphi}, \widetilde{\varphi'})$,
when expressed in terms of $\varphi$ and $\varphi'$, involves unpleasant kernel. This is however not so important
as the Hilbert space inner product plays the (important but) only technical role of controlling all the analytical subtleties.
It is the Krein inner product $(\varphi, \mathfrak{J''}\varphi')  = (\widetilde{\varphi}, \mathfrak{J'}
\widetilde{\varphi'})$ which serves to compute probabilities on the subspace of physical states
on which it is positive definite, and it is nice to have the relatively simple and explicit formula
 (\ref{krein-prod-1-photon}) for the Krein-inner product in the Krein space $(\mathcal{H''}, \mathfrak{J''})$
expressed in terms of position wave functions $\varphi, \varphi'$. 

We must be careful in preparing the fields as constructed with the hepl of white noise Hida operators. 
This can be achieved
by application of the general construction of Hida creation-annihilation operators to the nuclear space $\mathcal{S}_{0}(\mathbb{R}^3; \mathbb{C}^4)$ of restrictions of the Fourier transforms
of the test functions from $\mathcal{S}_{00}(\mathbb{R}^4; \mathbb{C}^4)$, to the positive energy sheet
${}_{\mathscr{O}_{\bar{p}}}$ of the ``light'' cone
in the momentum space. The point which requires a much analysis is the construction of the nuclear space
$\mathcal{S}_{0}(\mathbb{R}^3; \mathbb{C}^4)$ as a standard countably Hilbert space in the sense of Gelfand, which arises from a standard operator, which is necessary for te white noise construction of the electromegnetic potential field $A$:
\[
A(\varphi) = A^\mu(\varphi_\mu)= a(\widetilde{\varphi}|_{{}_{\mathscr{O}_{\bar{p}}}}) 
+ \eta a(\widetilde{\varphi}|_{{}_{\mathscr{O}_{\bar{p}}}})^+ \eta,
\]
where $a(\widetilde{\varphi}|_{{}_{\mathscr{O}_{\bar{p}}}}),a(\widetilde{\varphi}|_{{}_{\mathscr{O}_{\bar{p}}}})^+$ are the annihilation-creation generalized Hida operators, which for 
$\widetilde{\varphi}|_{{}_{\mathscr{O}_{\bar{p}}}} \in \mathcal{S}_{0}(\mathbb{R}^3; \mathbb{C}^4)$  coincide with
the  operators constructed by the application
of the Segal second quantization functor $\Gamma$ to the \emph{conjugated} {\L}opusza\'nski representation, $\eta$
is the Gupta-Bleuler operator equal $\Gamma(\mathfrak{J}')$,
and where $\varphi \in \mathcal{S}_{00}(\mathbb{R}^4)$, its Fourier transform $\widetilde{\varphi}$
belongs to $\mathcal{S}_{0}(\mathbb{R}^4)$ and where $\widetilde{\varphi} \mapsto 
\widetilde{\varphi}|_{{}_{\mathscr{O}_{\bar{p}}}}$ is the restriction to the cone, which turns out to be indeed 
a continuous map of nuclear spaces $\mathcal{S}_{0}(\mathbb{R}^4) \to \mathcal{S}_{0}(\mathbb{R}^3)$
(this requires a proof). 

It turns out that indeed the commutator
\[
[A(\varphi), A(\varphi')]
\] 
defines the kernel distribution equal to the Pauli-Jordan function multiplied by the minkowskian metric; 
in the proof one can apply e.g. the Kernel theorem as stated in \cite{Woronowicz}
(the ordinary Schwartz kernel theorem is not sufficient for the construction of the Wick product); 
and it follows that $A(\varphi)$ is the 
Wightman field transforming locally as a four vector field. The full construction of the free quantum 
electromagnetic field (and the quantum vector potential field) through the application of the Segal's functor $\Gamma$
of second quantization to the conjugate {\L}opusza\'nski representation we are going to publish in a separate
paper. 

It should be stressed that already the elements $\widetilde{\varphi}$ of the single particle space of the
{\L}opusza\'nski representation (and its conjugation) in the momentum picture do not in general fulfil 
the condition $p^\mu \widetilde{\varphi}_\mu = 0$, so that in general their Fourier transforms
$\varphi$ do not preserve the Lorentz condition $\partial^\mu \varphi_\mu = 0$. This corresponds to the well known
fact that the Lorentz condition cannot be preserved as an operator equation. It can be preserved in the sense of 
the Krein-product average on a subspace of Lorentz states which arise from the closed subspace
$\mathcal{H}_{\textrm{tr}}$ of the so called transversal states together with all their images under the action
of the {\L}opusza\'nski representation and its conjugation. We are now going to define the closed subspace
$\mathcal{H}_{\textrm{tr}}$.

\section{Construction of the physical subspace $\mathcal{H}_{\textrm{tr}}$ of transversal photon
states}\label{transversal}

Note that the operator $B$ of multiplication by the positive selfadjoint matrix (\ref{operatorB})
is selfadjoint in the Hilbert space $\oplus L^2(\mathbb{R}^3) = L^2(\mathbb{R}^3, \mathbb{C}^4)$ 
with respect to the ordinary invariant Lebesque measure
$\ud^3 \boldsymbol{\p}$ on $\mathbb{R}^3$ (the direct sum $\oplus$ is over the four components of the 
function $\widetilde{\varphi}$), and that the Hilbert space inner product in the single-photon 
state space $\mathcal{H}'$
is equal $(\cdot , \cdot) = (\cdot, B \cdot)_{{}_{L^2(\mathbb{R}^3, \mathbb{C}^4)}}$. The unitary operator
which has the direct integral decomposition
\begin{equation}\label{diagonalizingU}
\int \limits_{\mathscr{O}_{\bar{p}}} U_{\boldsymbol{\p}} \, \ud^3 \boldsymbol{\p}
\end{equation} 
(in the integral we use the spatial momentum coordinates $\boldsymbol{\p}$ on 
the cone $\mathscr{O}_{\bar{p}}$, and the integral may be treated as an integral on $\mathbb{R}^3$)
with each component $U_{\boldsymbol{\p}}$ being a unitary matrix operator in $\mathbb{C}^4$ transforming the standard basis
in $\mathbb{C}^4$ into the basis\footnote{Here $p \in \mathscr{O}_{\bar{p}}$ is regarded as the standard function of
spatial momentum coordinates $\boldsymbol{\p}$.} ${w_{{}_1}}^+(p), {w_{{}_1}}^-(p), w_{{}_{r^{-2}}}(p), w_{{}_{r^2}}(p)$
of eigenvectors of the hermitian matrix $B(p)$. It is easily seen that (\ref{diagonalizingU})
transforms the operator $B$, regarded as an operator in $L^2(\mathbb{R}^3, \mathbb{C}^4)$, into the orthogonal direct
sum of four multiplication operators on the measure space. Two first components of this direct sum are the multiplication
operators by the constant function equal to unity everywhere, the next direct summand is the operator of multiplication
by $\frac{1}{2}r^{-3}$ and the third orthogonal direct summand is the multiplication operator by $\frac{1}{2}r$
(recall that $r$ is the following function: $r(\boldsymbol{\p}) = | \boldsymbol{\p} | = 
\sqrt{\boldsymbol{\p} \cdot \boldsymbol{\p}}$). Therefore
the operator $B$ treated as an operator in $\mathcal{H}'$ is likewise unitarily equivalent to a 
direct sum of multiplication operators and thus self-adjoint. And similarly $B$ as the operator in 
$\mathcal{H}'$ has a pure point
spectrum $\{1\}$ consisting of just one element 1, and a continuous spectrum equal $\mathbb{R}_+$.
Indeed any element $\widetilde{\varphi} \in \mathcal{H}'$ may be uniquely written as the 
following linear combination 
\begin{equation}\label{eigenspaceB}
\widetilde{\varphi}(p) = {w_{{}_1}}^+(p) \, f_+(p) \, + \,
 {w_{{}_1}}^-(p) \, f_-(p) \,+ \, w_{{}_{r^{-2}}}(p)\, f_{0+}(p)
\, + \,  w_{{}_{r^2}}(p) \, f_{0-}(p)
\end{equation}
where $f_+, f_-, f_{0+}, f_{0-}$
are scalar functions on the light cone $\mathscr{O}_{\bar{p}}$. 
The first two functions $f_+, f_-$
run over the set of all square integrable functions on the light cone 
$\mathscr{O}_{\bar{p}}$ with respect to the invariant measure $\ud \mu |_{{}_{\mathscr{O}_{\bar{p}}}} = 
\frac{\ud^3 \boldsymbol{\p}}{| \boldsymbol{\p} |}$.
The functions $f_{0+}$ range over all functions on $\mathscr{O}_{\bar{p}}$ square integrable with respect to the measure 
$\frac{\ud^3 \boldsymbol{\p}}{| \boldsymbol{\p} |^3}$, and finally
 $f_{0-}$ range over all square integrable functions with respect to the measure\footnote{The measures 
$\frac{\ud^3 \boldsymbol{\p}}{| \boldsymbol{\p} |^3}$ and
$| \boldsymbol{\p} | \ud^3 \boldsymbol{\p}$ are of course not invariant on the cone, but note that the ordinary
Hilbert space inner product which they define on $\mathcal{H}'$ \emph{is not} the inner product preserved by 
the {\L}opusz\'nski representation. The representation preserves the Krein-inner product.}
$| \boldsymbol{\p} | \ud^3 \boldsymbol{\p}$.

Note that the four elements 
\[
{w_{{}_1}}^+ \, f_+, \,\,
 {w_{{}_1}}^- \, f_-,  \,\, w_{{}_{r^{-2}}} \, f_{0+},
\,\,  w_{{}_{r^2}} \, f_{0-}
\]
of $\mathcal{H}'$ on the right hand side of (\ref{eigenspaceB}) define orthogonal decomposition
$\mathcal{H}'$ into closed invariant subspaces of the self=adjoint operator $B$, treated as an operator in 
$\mathcal{H}'$. Moreover by the formula for the Krein inner-product in $\mathcal{H}', \mathfrak{J}'$
(compare (\ref{Kr-inn-Lop-1-space}) and (\ref{Kr-inn-Lop-1-space'})) the closed subspace spanned by the 
elements
\[
 w_{{}_{r^{-2}}} \, f_{0+},
\,\,  w_{{}_{r^2}} \, f_{0-},
\]
the closed subspace spanned by ${w_{{}_1}}^+ \, f_+, $ and te closed subspace spanned by
${w_{{}_1}}^- \, f_-$ are also mutually Krein-orthogonal.  

Let $\mathcal{H}_{\textrm{tr}}$ be the closed subspace of the Hilbert space $\mathcal{H}'$ corresponding
to the pure point spectrum $\{1\}$ of the operator $B$ in $\mathcal{H}'$. Then
$\mathcal{H}_{\textrm{tr}}$ is spanned by the elements 
\[
{w_{{}_1}}^+ \, f_+ \, + \,
 {w_{{}_1}}^- \,f_-
\] 
and the inner product of any two members of $\mathcal{H}_{\textrm{tr}}$ is equal to the Krein-inner product
which easily follows from the construction. 
Thus by construction for every element of $\mathcal{H}'$ existence and uniqueness of the projection on 
$\mathcal{H}_{\textrm{tr}}$ with respect to the Krein-inner product $(\cdot , \mathfrak{J}' \cdot)$ 
follows\footnote{Recall that for a general subspace 
in a Krein space neither the existence, nor the uniqueness of the projection of a vector  on the subspace
with respect to the Krein-inner-product is guaranteed. Thus its existence and uniqueness 
as well as the existence of the corresponding Krein-selfadjoint idempotent need to be proved.}.

It is important to understand that the properties of $\mathcal{H}_{\textrm{tr}}$ are of fundamental importance
for the construction of the physical space of transversal states and contrary to ordinary 
Hilbert space the stated above properties of the subspace $\mathcal{H}_{\textrm{tr}}$ are by far
not shared by a general (even closed) subspaces of a Krein space.

Because the inner product $(\cdot, \cdot)$ of $\mathcal{H}'$ is just equal to the positive inner product which
corresponds through the fundamental symmetry $\mathfrak{J}'$ to the Krein-inner product 
$(\cdot, \mathfrak{J}' \cdot)$ (in the notation
of \cite{Bog} $(\cdot, \cdot)_{\mathfrak{J}'} = (\cdot, \mathfrak{J}' \mathfrak{J}' \cdot)
= (\cdot, \cdot)$) it follows that the subspace $\mathcal{H}_{\textrm{tr}}$ is uniformly
positive in the sense of \cite{Bog}, V.5. Being a closed subspace $\mathcal{H}_{\textrm{tr}}$
is regular in the sense of \cite{Bog}, therefore by \cite{Bog}, Ch. V. the subspace
$\mathcal{H}_{\textrm{tr}}$ is orthocomplemented with respect to the Krein-inner-product
$(\cdot, \mathfrak{J}' \cdot)$ and admits unique projection on $\mathcal{H}_{\textrm{tr}}$
 with respect to the Krein-inner-product, which is bounded (boundedness, closedness, continuity always refer to the
ordinary Hilbert space inner product of $\mathcal{H}'$ or in general to the corresponding Hilbert space). 
Thus by \cite{Bog} there exist bounded Krein-selfadjoint idempotent $P$ (i.e. $P^2 = P$,
$P^\dagger = P$, where $P^\dagger = \mathfrak{J}' P^* \mathfrak{J}'$ with the ordinary adjoint $P^*$
in the Hilbert space $\mathcal{H}'$) with range $P\mathcal{H}' = \mathcal{H}_{\textrm{tr}}$. 

Now we define the elements of $\mathcal{H}_{\textrm{tr}}$ as the physical transversal states.
But it turns out that in order to account for the Lorentz covariance and the gauge freedom
we cannot stay within $\mathcal{H}_{\textrm{tr}}$. The {\L}opusza\'nski
representation and the representation conjugate to it, whenever applied to a vector $\widetilde{\varphi}$ of 
$\mathcal{H}_{\textrm{tr}}$, in general transform it into a vector $\widetilde{\varphi}''$
which does not lie in $\mathcal{H}_{\textrm{tr}}$. But the amazing property of these
representations is that always 
\begin{equation}\label{propertyLop}
\widetilde{\varphi}'' = \widetilde{\varphi}' + \widetilde{\varphi}_0
\end{equation}
for a unique vector $\widetilde{\varphi}' \in \mathcal{H}_{\textrm{tr}}$ and a unique 
$\widetilde{\varphi}_0$ whose Krein-inner-product norm vanishes:
\[
(\widetilde{\varphi}_0, \mathfrak{J}'\widetilde{\varphi}_0) = 0, \,\,\, 
\] 
where $(\cdot, \cdot)$ is the inner product in $\mathcal{H}'$,
and which is Krein-orthogonal to $\mathcal{H}_{\textrm{tr}}$: 
\[
(\widetilde{\varphi}_0, \mathfrak{J}' \widetilde{\varphi}''') = 0, \,\,\,
\widetilde{\varphi}''' \in \mathcal{H}_{\textrm{tr}}
\] 
(both $\widetilde{\varphi}'$ and $\widetilde{\varphi}_0$ in general depend on $\widetilde{\varphi}$
and on the applied transformation). Because the Krein-norm of 
$\widetilde{\varphi}'' = \widetilde{\varphi}' + \widetilde{\varphi}_0$
is equal to the Krein norm of $\widetilde{\varphi}'$, and the Krein inner product on 
$\mathcal{H}_{\textrm{tr}}$ coincides with the ordinary inner product on $\mathcal{H}'$,
and the representations are Krein-isometric, then it follows that the transformation
$\widetilde{\varphi} \mapsto \widetilde{\varphi}'$ which they generate on $\mathcal{H}_{\textrm{tr}}$
is isometric with respect to the ordinary Hilbert space-inner product induced on 
$\mathcal{H}_{\textrm{tr}}$ by the Krein inner product.

Moreover by construction of the dense core domain $\mathfrak{D}$ of the induced representation 
(compare \cite{wawrzycki-mackey}, Sect. 2) to which the {\L}opusza\'nski representation is equivalent
shows that $\mathfrak{D}$ is likewise dense in the subspace $\mathcal{H}_{\textrm{tr}}$.
It is easily seen because in our case $\mathfrak{D}$ consists of all those $\widetilde{\varphi} \in \mathcal{H}'$
which are continuous functions on the cone with compact support, and all of them when projected on 
$\mathcal{H}_{\textrm{tr}}$ include all the functions of the form
\[
{w_{{}_1}}^+ \, f_+ \, + \,
 {w_{{}_1}}^- \, f_-
\]
with $f_+, f_-$ continuous of compact support, which are obviously dense in $\mathcal{H}_{\textrm{tr}}$.
Therefore the representations generated by the action modulo unphysical states by the
 Krein representation (and its conjugation) on the transversal subspace $\mathcal{H}_{\textrm{tr}}$
is not only Hilbert-space isometric but can be uniquely extended to an ordinary unitary
representation on $\mathcal{H}_{\textrm{tr}}$. This is really amazing in view of the quite singular character
of the {\L}opusza\'nski representation (and its conjugation) for which representor of any boost
is unbounded (with respect to the Hilbert space norm of $\mathcal{H}'$). 
Now we show that the the {\L}opusza\'nski representation 
$WU^{{}_{(1,0,0,1)}{\L}}W^{-1}$ and its conjugation 
$\mathfrak{J}'WU^{{}_{(1,0,0,1)}{\L}}W^{-1}\mathfrak{J}'$ does have the property (\ref{propertyLop}). 
In fact during the proof we give explicit construction
of the unitary representation generated on the physical subspace $\mathcal{H}_{\textrm{tr}}$.

Indeed, in order to show (\ref{propertyLop}) it is sufficient to compute 
$\Lambda(\alpha^{-1}) w_{1}^{+}(\Lambda(\alpha)p)$ and $\Lambda(\alpha^{-1}) w_{1}^{-}(\Lambda(\alpha)p)$, 
for $\alpha \in SL(2,\mathbb{C})$.
It turns out that in general we have
\begin{equation}\label{Lambda-w}
\begin{split}
\Lambda(\alpha^{-1}) w_{1}^{+}(\Lambda(\alpha)p) = \Theta^{+}_{+}(\alpha, p) w_{1}^{+}(p) +  
\Theta^{+}_{-}(\alpha, p) w_{1}^{-}(p) + u^{+}(\alpha, p), \\
\Lambda(\alpha^{-1}) w_{1}^{-}(\Lambda(\alpha)p) = \Theta^{-}_{+}(\alpha, p) w_{1}^{+}(p) +  
\Theta^{-}_{-}(\alpha, p) w_{1}^{-}(p) + u^{-}(\alpha, p),
\end{split}
\end{equation}
where the scalar real functions $\Theta^{+}_{+}, \Theta^{+}_{-}, \Theta^{-}_{+}, \Theta^{-}_{+}$,
have the properties 
\[
 \Theta^{+}_{-} = -  \Theta^{-}_{+}, \,\, \Theta^{+}_{+} = \Theta^{-}_{-}
\]
and 
\[
(\Theta^{+}_{+})^2 + (\Theta^{-}_{+})^2 = 1;
\]
and where $u^{+}(\alpha, p)$ and $u^{-}(\alpha, p)$ are at ech point $p$ of the cone $\mathscr{O}_{\bar{p}}$ Krein-orthogonal to both $w_{1}^{+}(p)$ and $w_{1}^{-}(p)$ and have the Krein-norm zero in the Krein space
$\mathbb{C}^4, \mathfrak{J}_{\bar{p}}$:
\[
\begin{split}
\big(u^{+}(\alpha, p), \mathfrak{J}_{\bar{p}} w_{1}^{+}(p) \big)_{{}_{\mathbb{C}^4}} = 0,
\big(u^{+}(\alpha, p), \mathfrak{J}_{\bar{p}} w_{1}^{-}(p) \big)_{{}_{\mathbb{C}^4}} = 0, 
\big(u^{+}(\alpha, p), \mathfrak{J}_{\bar{p}} u^{+}(\alpha, p) \big)_{{}_{\mathbb{C}^4}} = 0, \\
\big(u^{-}(\alpha, p), \mathfrak{J}_{\bar{p}} w_{1}^{+}(p) \big)_{{}_{\mathbb{C}^4}} = 0,
\big(u^{-}(\alpha, p), \mathfrak{J}_{\bar{p}} w_{1}^{-}(p) \big)_{{}_{\mathbb{C}^4}} = 0,
\big(u^{-}(\alpha, p), \mathfrak{J}_{\bar{p}} u^{-}(\alpha, p) \big)_{{}_{\mathbb{C}^4}} = 0.
\end{split}
\]
In view of the properties (\ref{Kr-inn-Lop-1-space}) and (\ref{Kr-inn-Lop-1-space'}) of the Krein-inner product 
in the Krein space $(\mathcal{H}', \mathfrak{J}')$ of the representation 
$WU^{{}_{(1,0,0,1)}{\L}}W^{-1}$ and on account of the form (\ref{lop-rep-momentum}) of the representation, this implies in particular the property (\ref{propertyLop}) of the representation and in fact gives the concrete
shape of the unitary representation generated by its action modulo unphysical (of zero Krein norm
and zero projection on $\mathcal{H}_{\textrm{tr}}$) states on the closed subspace $\mathcal{H}_{\textrm{tr}}$.

Note that the action modulo unphysical states of the conjugate representation 
$\mathfrak{J}'WU^{{}_{(1,0,0,1)}{\L}}W^{-1}\mathfrak{J}'$ 
generates numerically identical unitary representation on $\mathcal{H}_{\textrm{tr}}$.
This is because the fundamental symmetry $\mathfrak{J}'$ acts as the identity operator
on $\mathcal{H}_{\textrm{tr}}$. Indeed because the decomposition components $\mathfrak{J}'_{p}$
of $\mathfrak{J}'$ in the direct integral decomposition of $\mathfrak{J}'$ have the form
(compare Appendix) $\mathfrak{J'}_p = V(\beta(p)^{-1}) \, \mathfrak{J}_{\bar{p}} V(\beta(p)) 
= \mathfrak{J}_{\bar{p}} V(\beta(p))^* \, V(\beta(p)) = \mathfrak{J}_{\bar{p}} B(p)$,
then we have
\[
\begin{split}
\mathfrak{J}'_p w_{1}^{+}(p) = w_{1}^{+},\\
\mathfrak{J}'_p w_{1}^{-}(p) = w_{1}^{-},\\
\mathfrak{J}'_p w_{r^{-2}}(p) = -w_{r^2},\\
\mathfrak{J}'_p w_{r^2}(p) = -w_{r^{-2}}.
\end{split}
\]
From this and from (\ref{Lambda-w}) it then easily follows
\begin{equation}\label{JLambdaJ-w'}
\begin{split}
\mathfrak{J'}_{{}_{p}} \Lambda(\alpha^{-1}) 
\mathfrak{J'}_{{}_{\Lambda(\alpha)p}} w_{1}^{+}(\Lambda(\alpha)p) = \Theta^{+}_{+}(\alpha, p) w_{1}^{+}(p) +  
\Theta^{+}_{-}(\alpha, p) w_{1}^{-}(p) + \mathfrak{J'}_{{}_{p}} u^{+}(\alpha, p), \\
\mathfrak{J'}_{{}_{p}} \Lambda(\alpha^{-1}) 
\mathfrak{J'}_{{}_{\Lambda(\alpha)p}} w_{1}^{-}(\Lambda(\alpha)p) = \Theta^{-}_{+}(\alpha, p) w_{1}^{+}(p) +  
\Theta^{-}_{-}(\alpha, p) w_{1}^{-}(p) + \mathfrak{J'}_{{}_{p}} u^{-}(\alpha, p).
\end{split}
\end{equation}
Now if $u \in \mathcal{H}_p = \mathbb{C}^4$ is any Krein null vector in 
$(\mathbb{C}^4, \mathfrak{J}_{\bar{p}})$, Krein orthogonal to $w_{1}^{+}(p)$ 
and $w_{1}^{-}(p)$ in $(\mathbb{C}^4, \mathfrak{J}_{\bar{p}})$ for every $p$ on the cone,
as is the case for both $u^{+}(\alpha, p), u^{-}(\alpha, p)$, then $u$ is of the form
\[
u = a w_{r^{-2}}^{+} + b w_{r^2}^{+}
\]
at each point $p$ (and $a,b$ depending on $p$). Because $w_{r^{-2}}^{+}, w_{r^2}^{+}$ are themselves
null in $(\mathbb{C}^4, \mathfrak{J}_{\bar{p}})$ for each $p$, then $u$ has to be of the form
\[
u = a w_{r^{-2}}^{+} \,\,\, \textrm{or} \,\,\, u = b w_{r^2}^{+}.
\]
Thus 
\[
\mathfrak{J}'_p u = -a w_{r^2} \,\,\, \textrm{or} \,\,\, 
\mathfrak{J}'_p u = -b w_{r^{-2}},
\]
and in any case the vector $\mathfrak{J}'_p u$ is likewise Krein-null and Krein orthogonal
to both $w_{1}^{+}, w_{1}^{-}$ at each point $p$ of the light cone in the Krein space
$(\mathbb{C}^4, \mathfrak{J}_{\bar{p}})$. Therefore we can rewrite (\ref{JLambdaJ-w'})
in the form
\begin{equation}\label{JLambdaJ-w}
\begin{split}
\mathfrak{J'}_{{}_{p}} \Lambda(\alpha^{-1}) 
\mathfrak{J'}_{{}_{\Lambda(\alpha)p}} w_{1}^{+}(\Lambda(\alpha)p) = \Theta^{+}_{+}(\alpha, p) w_{1}^{+}(p) +  
\Theta^{+}_{-}(\alpha, p) w_{1}^{-}(p) +  u^{+}_{{}_{*}}(\alpha, p), \\
\mathfrak{J'}_{{}_{p}} \Lambda(\alpha^{-1}) 
\mathfrak{J'}_{{}_{\Lambda(\alpha)p}} w_{1}^{-}(\Lambda(\alpha)p) = \Theta^{-}_{+}(\alpha, p) w_{1}^{+}(p) +  
\Theta^{-}_{-}(\alpha, p) w_{1}^{-}(p) +  u^{-}_{{}_{*}}(\alpha, p)
\end{split}
\end{equation}
where $u^{+}_{{}_{*}}, u^{-}_{{}_{*}}$ are Krein-null and Krein orthogonal
to both $w_{1}^{+}, w_{1}^{-}$ at each point $p$ of the light cone in the Krein space
$(\mathbb{C}^4, \mathfrak{J}_{\bar{p}})$. Thus our statement follows and the conjugate 
{\L}opusza\'nski representation  generates numerically the same unitary representation
on $\mathcal{H}_{\textrm{tr}}$ as does the {\L}opusza\'nski representation itself.
It is therefore sufficient to compute $\Theta^{+}_{+}, \Theta^{+}_{-}, \Theta^{-}_{+}, \Theta^{-}_{+}$
for the {\L}opusza\'nski representation as for the conjugate representation they are identical.

It is easily seen that the unitary representation
$\mathbb{U}, \mathbb{T}$ ($\mathbb{U}$ stands for representors
of the $SL(2, \mathbb{C})$ subgroup, and $\mathbb{T}$ stands for the representors of the translation subgroup) 
generated on $\mathcal{H}_{\textrm{tr}}$ has the form 
\[
\mathbb{U}(\alpha)\left( \begin{array}{c} f_{+} \\ 
                 f_{-} \end{array}\right)(p)
= \left( \begin{array}{cc} \Theta^{+}_{+}(\alpha, p) & \Theta^{-}_{+}(\alpha, p) \\ 
                          \Theta^{+}_{-}(\alpha, p) & \Theta^{-}_{-}(\alpha, p)  \end{array}\right)
\left( \begin{array}{c} f_{+}(\Lambda(\alpha) p) \\ 
                 f_{-}(\Lambda(\alpha) p) \end{array}\right),
\]
\[
\mathbb{T}(a)\left( \begin{array}{c} f_{+} \\ 
                 f_{-} \end{array}\right)(p)
= e^{ia\cdot p}
\left( \begin{array}{c} f_{+}(p) \\ 
                 f_{-}(p) \end{array}\right).
\]
In view of the indicated above properties of the functions $\Theta^{+}_{+}, \Theta^{+}_{-}, \Theta^{-}_{+}, 
\Theta^{-}_{+}$ there exists such a phase
$\Theta(\alpha, p)$ that $\Theta^{+}_{+}(\alpha, p) = \Theta^{-}_{-}(\alpha, p) = \cos \Theta(\alpha, p)$,
$\Theta^{-}_{+}(\alpha, p) = - \Theta^{+}_{-}(\alpha, p) =  \sin \Theta(\alpha, p)$, so that the unitary 
representors $\mathbb{U}(\alpha)$
of the subgroup $SL(2, \mathbb{C})$ of the unitary representation generated on $\mathcal{H}_{\textrm{tr}}$ 
may be written as 
\[
\mathbb{U}(\alpha)\left( \begin{array}{c} f_{+} \\ 
                 f_{-} \end{array}\right)(p)
= \left( \begin{array}{cc} \cos \Theta(\alpha, p) & \sin \Theta(\alpha, p) \\ 
                          -\sin \Theta(\alpha, p) & \cos \Theta(\alpha, p)  \end{array}\right)
\left( \begin{array}{c} f_{+}(\Lambda(\alpha) p) \\ 
                 f_{-}(\Lambda(\alpha) p) \end{array}\right),
\]
and recall that $(f_+, f_-)$ compose the Hilbert space of all pairs of functions on the cone
which are square integrable with respect to the invariant measure on the cone. Applying to this Hilbert space
and to the unitary representation $\mathbb{U},\mathbb{T}$ in $\mathcal{H}_{\textrm{tr}}$ (which is naturally identified
with the Hilbert space of pairs of functions $(f_+, f_-)$ square integrable on the light cone with the invariant measure) the unitary transformation $\mathcal{U}$ defined by 
\[
\mathcal{U}\left( \begin{array}{c} f_{1} \\ 
                 f_{-1} \end{array}\right)(p)
= \left( \begin{array}{cc} \frac{-i}{\sqrt{2}} & \frac{-i}{\sqrt{2}} \\ 
                          \frac{1}{\sqrt{2}} & \frac{-1}{\sqrt{2}}  \end{array}\right)
\left( \begin{array}{c} f_{+}(p) \\ 
                 f_{-}(p) \end{array}\right),
\]
we obtain
\begin{equation}\label{alpha-helicity+1-1}
\mathcal{U}^{-1}\mathbb{U}(\alpha) \mathcal{U} \left( \begin{array}{c} f_{1} \\ 
                 f_{-1} \end{array}\right)(p)
= \left( \begin{array}{cc}  e^{i \Theta(\alpha, p)} & 0 \\ 
                          0 &  e^{-i \Theta(\alpha, p)} \end{array}\right)
\left( \begin{array}{c} f_{1}(\Lambda(\alpha) p) \\ 
                 f_{-1}(\Lambda(\alpha) p) \end{array}\right),
\end{equation}
\begin{equation}\label{a-helicity+1-1}
\mathcal{U}^{-1} \mathbb{T}(a)\mathcal{U}\left( \begin{array}{c} f_{1} \\ 
                 f_{-1} \end{array}\right)(p)
= e^{ia\cdot p}
\left( \begin{array}{c} f_{1}(p) \\ 
                 f_{-1}(p) \end{array}\right),
\end{equation}
which is precisely the representation of the single-photon states described in 
\cite{bialynicki-2}, \S 4.3 (formulas (4.23) and (4.22)). For the connection of this
representation with the Riemann-Silberstein vector as well as for its construction 
based on the correspondence with energy-momentum of the classical field and the Staruszkiewicz affine connection
on the light cone \cite{star-cone}, we refer to \cite{bialynicki-2}.

Concerning the concrete form of the phase $\Theta$, or equivalently of the functions
$\Theta^{+}_{+}, \ldots \Theta^{-}_{-}$ in (\ref{Lambda-w}), it is essentially 
a matter of simple computation which follows from the rule (\ref{Lambda-w}) and the concrete
form of the eigenvectors ${w_{{}_1}}^+(p)$, ${w_{{}_1}}^-(p)$, $w_{{}_{r^{-2}}}(p)$, $w_{{}_{r^2}}(p)$.
Because the general formula is quite complicated and the computational recipe given here
is simple, we leave to the reader the computation of the general formula for $\Theta$. Here we give only
several examples, which nonetheless are sufficient for the proof of (\ref{Lambda-w}) and for the reconstruction
of the general formula for $\Theta$. To this end let $\alpha_{\mu\nu}$ denote the element of $SL(2, \mathbb{C})$
which corresponds through the natural homomorphism $\Lambda$ to the rotation
$\Lambda(\alpha_{\mu\nu})$ in the $\mu-\nu$ plane. In particular $\Lambda(\alpha_{03})$ denotes
the Lorentz rotation in the $0-3$ plane and $\Lambda(\alpha_{23})$ stands for the ordinary
spatial rotation in the $2-3$ plane, i.e. around the first axis. We agree to use
$\lambda$ to denote the hyperbolic angle of the Lorenz rotations $\Lambda(\alpha_{0k})$,
and let $\theta$ be the angle of the spatial rotations $\Lambda(\alpha_{ik})$.  
Then in particular we have    
\begin{multline*}
\Lambda({\alpha_{03}}^{-1}) w_{1}^{+}(\Lambda(\alpha_{03})p) = w_{1}^{+}(p), \\
\Lambda({\alpha_{03}}^{-1}) w_{1}^{-}(\Lambda(\alpha_{03})p) =  w_{1}^{-}(p) +  
\sinh \lambda \frac{\sqrt{(p^1)^2 + (p^2)^2}}{p^0 \cosh \lambda  + p^3 \sinh \lambda} \, w_{{}_{r^{-2}}};
\end{multline*}
\[
\begin{split}
\Lambda({\alpha_{12}}^{-1}) w_{1}^{+}(\Lambda(\alpha_{12})p) = w_{1}^{+}(p), \\
\Lambda({\alpha_{12}}^{-1}) w_{1}^{-}(\Lambda(\alpha_{12})p) =  w_{1}^{-}(p);
\end{split}
\]
\begin{multline*}
\Lambda({\alpha_{23}}^{-1}) w_{1}^{+}(\Lambda(\alpha_{23})p) = \\
\frac{1}{\sqrt{(p^1)^2 + (p^2 \cos \theta + p^3 \sin \theta)^2}} 
\Big( \frac{p^2 p^3 \sin \theta}{\sqrt{(p^1)^2 + (p^2)^2}} + \sqrt{(p^1)^2 + (p^2)^2} \cos \theta \Big) w_{1}^{+}(p) \\
+ \frac{r p^1 \sin \theta}{\sqrt{(p^1)^2 + (p^2 \cos \theta + p^3 \sin \theta)^2} \, \sqrt{(p^1)^2 + (p^2)^2}}
 w_{1}^{-}(p), 
\end{multline*}

\begin{multline*}
\Lambda({\alpha_{23}}^{-1}) w_{1}^{-}(\Lambda(\alpha_{23})p) = \\ 
\frac{-r p^1 \sin \theta}{\sqrt{(p^1)^2 + (p^2 \cos \theta + p^3 \sin \theta)^2} \, \sqrt{(p^1)^2 + (p^2)^2}}
 w_{1}^{+}(p) \\
+ \frac{1}{\sqrt{(p^1)^2 + (p^2 \cos \theta + p^3 \sin \theta)^2}} 
\Big( \frac{p^2 p^3 \sin \theta}{\sqrt{(p^1)^2 + (p^2)^2}} + \sqrt{(p^1)^2 + (p^2)^2} \cos \theta \Big) w_{1}^{-}(p).
\end{multline*}
The formulas for $\Lambda({\alpha_{13}}^{-1}) w_{1}^{+}(\Lambda(\alpha_{13})p)$ and 
$\Lambda({\alpha_{13}}^{-1}) w_{1}^{-}(\Lambda(\alpha_{13})p)$ are obtained 
by interchanig $p^1$ and $p^2$ with each other in the last two formulas respectively.

From the the first two of the formulas just written it is easily seen that the representors of the Lorentz
rotations $\Lambda(\alpha_{03})$ in the {\L}opusza\'nski representation are unbounded, as their action on any element
\[
{w_{{}_1}}^+ \, f_+ \, + \,
 {w_{{}_1}}^- \,f_-
\] 
of $\mathcal{H}_{\textrm{tr}}$ equals 
\[
\begin{split}
p \mapsto {w_{{}_1}}^+(p) \, f_+(\Lambda(\alpha_{03})p) \, + \,
 {w_{{}_1}}^-(p) \,f_-(\Lambda(\alpha_{03})p) \\
+ \sinh \lambda \frac{\sqrt{(p^1)^2 + (p^2)^2}}{p^0 \cosh \lambda + p^3 \sinh \lambda} \, 
w_{{}_{r^{-2}}}(p) \, f_{-}(\Lambda(\alpha_{03})p);
\end{split}
\]
and on the other hand the function $p \mapsto f_-(p)$ being square integrable with respect to the invariant 
measure $r^{-1}\ud^3 \boldsymbol{\p}$ gives the function $p \mapsto f_-(\Lambda(\alpha_{03})p)$
which is likewise square integrable with respect to the invariant measure on the cone. But in general
such a function is not square integrable with respect to the measure $r^{-3}\ud^3 \boldsymbol{\p}$
on the cone.

In general only the Lorentz rotations (``boosts'') produce the additional unphysical Krein zero vector $u$ Krein  
orthogonal to $\mathcal{H}_{\textrm{tr}}$ when acting on $\mathcal{H}_{\textrm{tr}}$. Spatial rotations
(and of course translations) transform unitarily $\mathcal{H}_{\textrm{tr}}$ into itself. 

From the above formulas it follows for example that 
\[
\begin{split}
\sin \Theta(\alpha_{03},p) = 0, \, \cos \Theta(\alpha_{03},p) = 1, \\
\sin \Theta(\alpha_{12},p) = 0, \, \cos \Theta(\alpha_{12},p) = 1, \\
\sin \Theta(\alpha_{23},p) = 
\frac{r p^1 \sin \theta}{\sqrt{(p^1)^2 + (p^2 \cos \theta + p^3 \sin \theta)^2} \, \sqrt{(p^1)^2 + (p^2)^2}}; \\
 \cos \Theta(\alpha_{23},p) =
\frac{1}{\sqrt{(p^1)^2 + (p^2 \cos \theta + p^3 \sin \theta)^2}} 
\Big( \frac{p^2 p^3 \sin \theta}{\sqrt{(p^1)^2 + (p^2)^2}} + \sqrt{(p^1)^2 + (p^2)^2} \cos \theta \Big).
\end{split}
\] 
The formulas for $\sin \Theta(\alpha_{13},p)$ and 
$\cos \Theta(\alpha_{13},p)$ are obtained 
by interchanig $p^1$ and $p^2$ with each other in the last two formulas respectively.

The representation given by (\ref{alpha-helicity+1-1}) and (\ref{a-helicity+1-1}) has already the 
form of direct sum of unitary
representations concentrated on the single (zero mass) orbit 
$\mathscr{O}_{(1,0,0,1)} \cong SL(2, \mathbb{C}) /G_{(1,0,0,1)}$, with the opposite signs of helicities
of the representations of the small group $G_{(1,0,0,1)}$ (double cover of the Euclidean group). 
Because $\Theta$ depends on $\alpha \in SL(2, \mathbb{C})$ only through the natural
homomorphism $\Lambda$ of the $SL(2, \mathbb{C})$ onto the Lorentz group, then it is easily
seen that the helicities must be integer numbers (and not half-integer), and thus giving
true (single valued) representations of the Poincar\'e group (and not the double valued). A simple computation
shows that (\ref{alpha-helicity+1-1}) and (\ref{a-helicity+1-1}) is unitarily equivalent to the direct sum of unitary representations
induced respectively by the irreducible helicity 1 and -1 representations of the small subgroup $G_{(1,0,0,1)}$.
Thus according to the Mackey theory of induced representations (\ref{alpha-helicity+1-1}) and (\ref{a-helicity+1-1}) 
is equivalent to $[m=0, h=1] \oplus [m=0, h=-1]$.

Apparently we could restrict to the subspace  $\mathcal{H}_{\textrm{tr}}$ and to the 
unitary representation (\ref{alpha-helicity+1-1}) and (\ref{a-helicity+1-1})
acting in $\mathcal{H}_{\textrm{tr}}$, forgetting about the surrounding Krein space and the {\L}opusza\'nski
represntation and its conjugation. But although the representation (\ref{alpha-helicity+1-1}) and (\ref{a-helicity+1-1})
is localizable in the sense of Jauch and Piron \cite{jauch-piron} (for the proof compare \cite{amrein})
it is not equivalent to any representation which in the momentum picture has the multiplier independent
of $p \in \mathscr{O}_{\bar{p}}$, i.e. the transformation it generates in the position picture is nonlocal. 
It is really amazing that in passing from the vector potential to the Riemann-Silberstein vector (or to the electric and
magnetic field) the $p$-dependence of the multplier of the representation (\ref{alpha-helicity+1-1}) 
and (\ref{a-helicity+1-1}) is compensated for by the $p$-dependence of the polarization vector, so that the application
of the second quantization functor $\Gamma$ to the representation (\ref{alpha-helicity+1-1}) and (\ref{a-helicity+1-1})
gives indeed the quantum Riemann-Silberstein (or quantum electric and magnetic) field with local transformation formula,
compare \cite{bialynicki-2}. Unfortunately the compensation breaks down for the vector potential itself
(the additional factor $p^0 (\boldsymbol{\p})$ is crucial when passing form the pair of helicity +1 and -1 
vector potentials to the Riemann-Silberstein vector) so that 
locality of the vector potential is not preserved within this scheme.
We need the surrounding Krein space and the {\L}opusza\'nski representation (and its conjugation) in 
order to save the local character of the transformation law for the vector potential. Alternatively
one can interprete the situation as follows. In order to restore the locality of the inverese Fourier transforms of the transformation of the elements
${w_{{}_1}}^+ \, f_+ \, + \, {w_{{}_1}}^- \,f_-$ of $\mathcal{H}_{\textrm{tr}}$ acted on by the representation $(\mathbb{U}, \mathbb{T})$ one has to additionally
apply the appropriate gauge transformation, i.e. one has to add to the inverse Fourier transforms of the transformed element
$\mathbb{U}(\alpha) \big( {w_{{}_1}}^+ \, f_+ \, + \, {w_{{}_1}}^- \,f_- \big)$ a gradient
of a function $g$ depending on the initial state (e.g. on the inverse Fourier transform of ${w_{{}_1}}^+ \, f_+ \, + \, {w_{{}_1}}^- \,f_-$) highly non-locally, and explicitly on $x$ and $\alpha$. Accorningly to the reseults presented above, the 
$p$-dependence of the phase $\Theta$ is just compensated for by the $p$ dependence of the eigenvectors  $w_{1}^{+},
w_{1}^{-}$, whenever $\alpha$ corresponds to the spatial rotation,
so that the transformation remains local in position picture in this particular case.
But for a general Lorentz rotation, $u^{+}, u^{-}$ (and $\Theta$) depend on $p$ nontrivially so that this dependence is not canclelled by the $p$-dependence of $w_{1}^{+}, w_{1}^{-}$ and $\mathbb{U}(\alpha)$ in the position picture acts in general non-locally. Indeed, in general the difference between the representations $U$ and $\mathbb{U}$ acting on $\mathcal{H}_{\textrm{tr}}$ is equal
\[
U(\alpha) \big( {w_{{}_1}}^+ \, f_+ \, + \, {w_{{}_1}}^- \,f_- \big) - \mathbb{U}(\alpha) \big( {w_{{}_1}}^+ \, f_+ \, + \, {w_{{}_1}}^- \,f_- \big) =
\left( \begin{array}{c} p^0 \\
                              p^1    \\
                           p^2    \\
                             p^3       \end{array}\right) \, \cdot \widetilde{g}(\alpha, f_+, f_-, p),
\]
where the scalar function $\widetilde{g}$ depends on $\alpha, , f_+, f_-, p$. Therefore in the position picture
the difference between the local transformatin $U(\alpha)$ and the nonlocal transformation 
$\mathbb{U}(\alpha)$ is equal to a gradnient of a function $g$, depending on $\alpha, f_+, f_-$
and explicitly on $x$. The fumction $U(\alpha) \big( {w_{{}_1}}^+ \, f_+ \, + \, {w_{{}_1}}^- \,f_- \big)$ belongs to 
$\mathcal{H}'$ whenewer $\widetilde{g}$ is square integrable on the cone with respect to the invariant measure.  In particlular
\[
\widetilde{g}(\alpha_{03}, f_+, f_-, p) = 
\sinh \lambda \frac{\sqrt{(p^1)^2 + (p^2)^2}}{p^0 \cosh \lambda + p^3 \sinh \lambda} \, 
\frac{1}{\sqrt{2} p^0} \, f_{-}(\Lambda(\alpha_{03})p);
\]   
so that it is easily seen that in passing to the position picture the inverse Fourier transform
$g$ of $\widetilde{g}$ depend in a non-local way on the initial state (inverese Fourier transform of 
${w_{{}_1}}^+ \, f_+ \, + \, {w_{{}_1}}^- \,f_-$).

\subsection{\bf{Appendix}: Krein-isometric representations concentrated on single orbits 
and local wave functions}\label{constr-of-VF}

This Section is based on
a generalization of Mackey'a theory, presented in \cite{wawrzycki-mackey}.

We assume the results of the mentioned generalization and use them 
in the construction of 
single particle wave functions which in the position picture have local transformation rule.
The novelty lies in the application to Krein-isometric representations in Krein spaces and in that it 
is more consequently related to the theory of induced representations. 

Representations of the double cover $T_{4} \circledS SL(2, \mathbb{C})$ of the 
Poincar\'e group\footnote{We denote the representor of $(a,\alpha) \in T_{4} \circledS SL(2, \mathbb{C})$
by $U_{(a,\alpha)}$, and the convention in which the Lorentz transformation $\Lambda(\alpha)$ corresponding
to $\alpha \in SL(2, \mathbb{C})$ is an antihomomorphism, and with the right action of $\Lambda$ on $a \in
\widehat{T_4}$.}
considered here are in general not unitary but Krein-unitary and even
only Krein-isometric (for definitions compare Sect. 1 and 2 of \cite{wawrzycki-mackey})
with the properties (\ref{circumstance}) and (\ref{circumstance2}) motivated by the 
properties of representations acting in the Krein-Fock spaces
of the free fields underlying QED (and the Standard Model).  The first property is that the 
Gupta-Bleuler  operator $\mathfrak{J}$ -- plying the role of the fundamental symmetry of the Krein
space in question, commutes with translations. Consider first
a Krein-isometric representation acting in one particle Krein subspace $(\mathcal{H}, \mathfrak{J})$ 
(or in its subspace) of the Krein-Fock space in question. Because translations (we mean of course their representors)
commute with $\mathfrak{J}$, they are not only Krein-isometric but unitary with respect to the Hilbert space inner product
of the Krein space $(\mathcal{H}, \mathfrak{J})$ in question. 
Let $P^0 , \ldots , P^3$ be the respective generators of the translations (they do exist by the strong continuity assumption posed on the Krein-isometric representation -- physicist's everyday computations involve the generators 
and thus our assumption is justified, compare Sect. 2 of \cite{wawrzycki-mackey}).  Let $\mathcal{C}$ be the commutative $C^*$-algebra generated 
by the functions $f(P^0 , \ldots ,P^3)$ of translation generators $P^0 , \ldots , P^3$,
where $f$ is continuous on $\mathbb{R}^4$ and vanishes at infinity. Let 
\begin{equation}\label{dec-sp-P}
\mathcal{H} = \int \limits_{sp(P^0 , \ldots , P^3)} \mathcal{H}_{p} \, \ud \mu (p)
\end{equation} 
be the direct integral decomposition of $\mathcal{H}$ corresponding to the algebra $\mathcal{C}$
(in the sense of \cite{von_neumann_dec} or \cite{Segal_dec_I}) with a spectral measure
$\mu$ on the joint spectrum sp($P^0 , \ldots , P^3$) of the translation generators.
We may identify sp($P^0 , \ldots , P^3$) with a subset of the group $\widehat{T_4}$ dual to  the translation group
$T_4$. Moreover we may assume that the algebra $\mathcal{C}$ and the spectral measure corresponding to the
above decomposition (\ref{dec-sp-P}) are of uniform multiplicity, compare Theorem 5, Sect. 5 of 
\cite{wawrzycki-mackey}.
Let us denote the translation representor $U_{(a,1)}$ just by $T(a)$ and the representor $U_{(0,\alpha)}$ of the 
$SL(2, \mathbb{C})$ subgroup just by $U(\alpha)$. 
By the multiplication rule in $T_{4} \circledS SL(2, \mathbb{C})$ it follows that 
\begin{equation}\label{utu}
T(a\Lambda(\alpha^{-1})) = U(\alpha)^{-1} T(a) U(\alpha), 
\end{equation}
such that
\[
U(\alpha)^{-1} P^{\nu} U(\alpha) = \Lambda(\alpha^{-1})_{\mu}^{\nu} P^{\mu}  \,\,\,(\textrm{summation \, over} \,\, \mu)
\]
so that $U(\alpha)^{-1} E(S) U(\alpha) = E(\Lambda(\alpha^{-1})S)$ for $S \subset{}$ sp($P^0 , \ldots , P^3$),
i. e. $U(\alpha)$ acts on the joint spectrum of $P^0 , \ldots , P^3$ as the ordinary right action
of the Lorentz transformation $\Lambda(\alpha^{-1})$.  
Moreover we may identify 

sp$(P^0 , \ldots , P^3) \subset \widehat{T_4}$ with the orbit $\mathscr{O}_{\bar{p}}$ under the standard action of the Lorentz group of a single point $\bar{p} = \bar{p}(m)$ in the vector space $\mathbb{R}^4$ endowed with the
 Minkowski pseudo-metric form $g_M$ with the signature $(1, -1, -1, -1)$, and with the invariant measure
$\mu_{m}$ on the orbit $\mathscr{O}_{\bar{p}} = \{p: g_M (p , p) = m^2\}$ induced by the invariant Lebesgue measure on 
$\mathbb{R}^4$ equal to the Haar measure on $\widehat{T_4}$.  
Because the fundamental symmetry $\mathfrak{J}$ commutes with $P^0 , \ldots , P^3$ it is decomposable
with respect to the decomposition (\ref{dec-sp-P}), and let $p \mapsto \mathfrak{J}_p$ be its
decomposition with respect to (\ref{dec-sp-P}), i. e.
\[
\mathfrak{J} = \int \limits_{sp(P^0 , \ldots , P^3) \cong \mathscr{O}_{\bar{p}}} \mathfrak{J}_p \, 
\ud \mu |_{{}_{\mathscr{O}_{\bar{p}}}} (p)
\]
with $\mathfrak{J}_p$ being a fundamental symmetry in $\mathcal{H}_p$. Because of the uniform multiplicity
$\mathcal{H}_p \cong \mathcal{H}_{\bar{pwig}}$, $p \in \mathscr{O}_{\bar{p}}$. Moreover every element
$\widetilde{\psi} \in \mathcal{H}$ may be identified with the function 
$\mathscr{O}_{\bar{p}} \ni p \mapsto \widetilde{\psi}(p) \in \mathcal{H}_{\bar{p}}$
equal to the decomposition of $\widetilde{\psi}$ with respect to (\ref{dec-sp-P}). 
Therefore in the notation of von Neumann 
\[
\widetilde{\psi} = \int \limits_{sp(P^0 , \ldots , P^3) \cong \mathscr{O}_{\bar{p}}}  
\widetilde{\psi}(p) \, \sqrt{\ud \mu |_{{}_{\mathscr{O}_{\bar{p}}}} (p)}.
\]

We may assume that 
$\mathfrak{J}_p$ does not depend on $p$ -- which is still sufficient for the representations acting
on one-particle states as well as for the decomposition of their tensor products (the latter assertion
will be proved in the further stages of this paper). In this Section we identify
every $\widetilde{\psi} \in \mathcal{H}$ with the corresponding function $p \mapsto \widetilde{\psi}(p)$
-- its decomposition.

Now for each $\alpha \in SL(2, \mathbb{C})$ let us define the following operator $D(\alpha)$ (compare
\cite{Ohnuki, Wigner_Poincare})
\[
D(\alpha) \widetilde{\psi}(p) = \widetilde{\psi}(\Lambda(\alpha)p).
\] 
By the Lorentz invariance of the measure $\mu$ on the orbit $\mathscr{O}_{\bar{p}}$ it follows that
$D(\alpha)$ is unitary for every $\alpha \in SL(2, \mathbb{C})$. Moreover, because the components 
$\mathfrak{J}_p$ in the decomposition of $\mathfrak{J}$ do not depend on $p \in \mathscr{O}_{\bar{p}}$,
it easily follows that $D(\alpha)$ commutes with $\mathfrak{J}$, so that $D(\alpha)$ is Krein-unitary
for each $\alpha \in SL(2, \mathbb{C})$. Thus $\alpha \mapsto D(\alpha)$ gives a unitary and Krein-unitary representation
of $SL(2, \mathbb{C})$:
\[
D(\alpha \beta) = D(\alpha) D(\beta).
\]

Let $F$ be any Baire function on sp$(P^0 , \ldots , P^3) = \mathscr{O}_{\bar{p}}$, and let 
$F(P) = F(P^0 , \ldots , P^3)$ be the operator function of $P^0 , \ldots , P^3$, i. e. operator 
\[
F(P)\widetilde{\psi}(p) = F(p) \widetilde{\psi}(p).
\]
An easy computation shows that
\begin{equation}\label{[D,F]}
D(\alpha) \, F(P) = F\big( \Lambda(\alpha)P \big) \, D(\alpha),
\end{equation}
where $F(\Lambda(\alpha)P) = F(\Lambda(\alpha)_\nu^\mu P^\nu)$ (summation with respect to $\nu$).
Joining (\ref{utu}) and (\ref{[D,F]}) it follows that 
\begin{equation}\label{[UD^-1,T]}
\big[ U(\alpha)D(\alpha)^{-1}, T(a) \big] = 0.
\end{equation}
Thus $Q(\alpha) = U(\alpha)D(\alpha)^{-1}$ commutes with the elements of the $C^*$- algebra $\mathcal{C}$ 
and it is decomposable with respect to (\ref{dec-sp-P}) (in other words it is a function of the 
operators $P^0 , \ldots , P^3$). Denote the components $Q(\alpha)_p$ of $Q(\alpha)$ with respect to
this decomposition just by $Q(\alpha , p)$. Recall that they are operators acting in $\mathcal{H}_{\bar{p}}$,
so that 
\[
Q(\alpha) = \int \limits_{\mathscr{O}_{\bar{p}}} Q(\alpha , p) \, \ud \mu |_{{}_{\mathscr{O}_{\bar{p}}}} (p).
\]
Thus in the notation of von Neumann \cite{von_neumann_dec} 
\[
U(\alpha) \widetilde{\psi} = Q(\alpha) D(\alpha) \widetilde{\psi}
= \int \limits \limits_{\mathscr{O}_{\bar{p}}} Q(\alpha , p) \big(D(\alpha)\widetilde{\psi} \big)(p) \, 
\sqrt{\ud \mu |_{{}_{\mathscr{O}_{\bar{p}}}} (p)},
\]  
where $p \mapsto  \big(D(\alpha)\widetilde{\psi} \big)(p)$ is the decomposition of  $D(\alpha)\widetilde{\psi}$,
so that 
\[
p \mapsto \Big( U(\alpha) \widetilde{\psi} \Big)(p) =  Q(\alpha , p) \big(D(\alpha)\widetilde{\psi}\big)(p)
\]
is the decomposition of $U(\alpha) \widetilde{\psi}$.

Because $\alpha \mapsto U(\alpha)$ is a representation it follows that the components $Q(\alpha , p)$ of
$Q(\alpha)$ have the following multiplier property
\[
Q(\delta \alpha), p) = Q(\delta, p) Q(\alpha, \Lambda(\delta)p), \,\,\, p \in \mathscr{O}_{\bar{p}}, 
\alpha,\delta \in SL(2, \mathbb{C}).
\]
In particular
\[
Q(e,p) = 1 , \,\,\, Q(\alpha, p)^{-1} = Q(\alpha^{-1}, \Lambda(\alpha)p).
\]
If we consider any Krein-isometric operator $W$ which preserves the invariant core domain of the Krein-isometric
representation $U$ (i.e. the domain $\mathfrak{D}$ of Sect. 2 of \cite{wawrzycki-mackey}) and which is decomposable
with respect to (\ref{dec-sp-P}) with the decomposition $p \mapsto W(p)$, then (with $\widetilde{\Psi }= 
W\widetilde{\psi}$)
\begin{equation}\label{WUW^-1}
WU(\alpha)W^{-1} \widetilde{\Psi} = \int \limits_{\mathscr{O}_{\bar{p}}} 
W(p) Q(\alpha, p) W(\Lambda(\alpha)p)^{-1} \big(D(\alpha)\widetilde{\Psi}\big)(p) \, 
\sqrt{\ud \mu |_{{}_{\mathscr{O}_{\bar{p}}}} (p)}
\end{equation}
with $WU(\alpha)W^{-1}$ being another Krein-isometric representation, forces 
\begin{equation}\label{Q'}
Q'(\alpha, p)= W(p) Q(\alpha) W(\Lambda(\alpha)p)^{-1}
\end{equation}
to be another multiplier:
\[
Q'(\delta \alpha), p) = Q'(\delta, p) Q'(\alpha, \Lambda(\delta)p), \,\,\, p \in \mathscr{O}_{\bar{p}}, 
\alpha,\beta \in SL(2, \mathbb{C}),
\]
corresponding to the representation $\alpha \mapsto WU(\alpha)W^{-1}$.

  Moreover the core domain $\mathfrak{D}$ have the following \emph{pervasive}\footnote{Term introduced by 
Mackey in \cite{Mackey}.} property that there exist a sequence $\{f_l\}_{l \in \mathbb{N}}$
of elements of $\mathfrak{D}$ such that for all $p \in {}$ sp$(P^0 \ldots , P^3) = \mathscr{O}_{\bar{p}}$
(compare Lemma 6 of \cite{wawrzycki-mackey}) $\{f_l (p)\}_{l \in \mathbb{N}}$ is dense in $\mathcal{H}_{\bar{p}} = \mathcal{H}_p$.

Now the operator $Q(\alpha,p)$ is Krein-unitary for almost all $p \in \mathscr{O}_{\bar{p}}$.
Indeed we have
\begin{multline*}
Q(\alpha, p) \mathfrak{J}_{\bar{p}}Q(\alpha, p)^*\mathfrak{J}_{\bar{p}} f_{l} (p) = f_l (p) \,\,\, \textrm{and} \\
\mathfrak{J}_{\bar{p}}Q(\alpha, p)^*\mathfrak{J}_{\bar{p}} f_{l} (p) Q(\alpha, p) = f_l (p) 
\,\,\, p \in \mathscr{O}_{\bar{p}}, l \in \mathbb{N}.
\end{multline*}
Because for each $p \in \mathscr{O}_{\bar{p}}$, $\{f_l (p)\}_{l \in \mathbb{N}}$ is dense in $\mathcal{H}_{\bar{p}}$
and because the representation $\alpha \mapsto U(\alpha)$ is locally finite with respect to the spectral measure 
$E$ of $T$ determining the corresponding direct integral decomposition (\ref{dec-sp-P}), i.e. fulfils 
(\ref{circumstance2}), then 
\[
Q(\alpha, p) \mathfrak{J}_{\bar{p}} Q(\alpha,p)^*\mathfrak{J}_{\bar{p}} = \bold{1} \,\,\, \textrm{and} \,\,\, \\
\mathfrak{J}_{\bar{p}} Q(\alpha,p)^*\mathfrak{J}_{\bar{p}} Q(\alpha,p) = \bold{1},
\] 
and $Q(\alpha, p)$ is Krein-unitary for all $p \in \mathscr{O}_{\bar{p}}$. In fact in case of the single particle representations the restriction $T$ of the representation to translations has finite uniform multiplicity
so that $\mathcal{H}_{\bar{p}}$ has finite dimension, so that the unitarity of $Q(\alpha, p)$
for almost all $p$ immediately follows independently of the assumption (\ref{circumstance2}) in this case. 

It is well known that each element $p = (p^0 , \ldots , p^3) \in \mathbb{R}^4$ of the dual group
$\widehat{T_4} \supset {}$ sp$(P^0 , \ldots , P^3)$ may be represented by the hermitean $2 \times 2$ matrix 
$\hat{p} = p^0 \bold{1}  + p^1 \sigma_1 + p^2 \sigma_2 + p^3 \sigma_3$, where $\sigma_i$
are the Pauli matrices, and with the action of the Lorentz transformation $\Lambda(\alpha)p$ on
$p$ given by $\alpha \hat{p}\alpha^* = \widehat{\Lambda(\alpha^{-1})p}$. Now let $\bar{p}$ be any fixed
point of the orbit $\mathscr{O}_{\bar{p}}$. Now we associate bi-uniquely an element 
$\beta(p) \in SL(2, \mathbb{C})$ with every $p \in \mathscr{O}_{\bar{p}}$ such that 
$\beta(p)^{-1} \widehat{\bar{p}}{\beta(p)^*}^{-1} = \hat{p}$, i. e. $\Lambda(\beta(p))\bar{p} = p$
and $\Lambda(\beta(p)^{-1})p = \bar{p}$. Of course the function $p \mapsto \beta(p) = \beta_{\bar{p}}(p)$ depends
on the orbit $\mathscr{O}_{\bar{p}}$, but we discard the subscript $\bar{p}$ at $\beta(p)$ in order to simplify
notation, as in the most part of this Sect. we are concerned with a fixed orbit. 

It follows that $\gamma(\alpha,p) = \beta(p) \alpha \beta(\Lambda(\alpha)p)^{-1}$ is an element of the subgroup
$G_{\bar{p}}$ stationary for $\bar{p}$: $\Lambda(\gamma(\alpha,p))\bar{p} = \bar{p}$, or 
$\gamma(\alpha,p)\widehat{\bar{p}}\gamma(\alpha,p)^{*} = \widehat{\bar{p}}$. Therefore
every $\alpha \in SL(2,\mathbb{C})$ has the following factorization:
\[
\alpha = \beta(p)^{-1} \gamma(\alpha,p)\beta(\Lambda(\alpha)p).
\]

Thus because $Q(\alpha,p)$ is a multiplier we obtain
\begin{multline}\label{WQW^-1}
Q(\alpha, p) = Q\big(\beta(p)^{-1} \gamma(\alpha,p) \beta(\Lambda(\alpha)p, p \big) \\
= Q(\beta(p)^{-1},p) \, Q(\gamma(\alpha, p), \bar{p}) \, Q(\beta(\Lambda(\alpha)p), \bar{p}).
\end{multline}

Now let us introduce the operator $W$ decomposable with respect to (\ref{dec-sp-P})
whose decomposition function is given by 
\begin{equation}\label{W1}
p \mapsto W(p) =  Q(\beta(p),\bar{p}).
\end{equation}
Because the components $Q(\alpha,p)$ of $Q(\alpha)$ compose a multiplier, then
$W(p)^{-1} = Q(\beta(p)^{-1},p)$, so that the operator $W^{-1}$ has the
decomposition $p \mapsto W(p)^{-1} = Q(\beta(p)^{-1},p)$. By construction $W$ is a Krein-isometric
operator which preserves the core domain $\mathfrak{D}$ of the initial representation $U$ and
moreover by (\ref{WQW^-1}) we have:
\[
Q(\alpha, p) = W(p)^{-1} Q(\gamma(\alpha,p), \bar{p}) 
W(\Lambda(\alpha)p).
\] 
Comparison with (\ref{WUW^-1}) and (\ref{Q'}) shows that the original Krein-isometric representation
$U$ is equivalent to the Krein-isometric representation $W^{-1}UW$, where\footnote{We have denoted 
$\widetilde{\psi}$ and $W\widetilde{\psi}$ by the same letter $\widetilde{\psi}$, we hope this will not
cause any misunderstanding.}
\[
p \mapsto W^{-1}U(\alpha)W \widetilde{\psi} (p) 
= Q(\gamma(\alpha,p), \bar{p}) D(\alpha)\widetilde{\psi} (p)
=  Q(\gamma(\alpha,p), \bar{p}) \widetilde{\psi} (\Lambda(\alpha)p)
\]
is the decomposition of $W^{-1}U(\alpha)W \widetilde{\psi}$. Note that for $\gamma, \gamma'$
ranging over the subgroup $G_{\bar{p}}$ stationary for $\bar{p}$ we have
\[
Q(\gamma \gamma',\bar{p}) = Q(\gamma, \bar{p})Q(\gamma', \bar{p}),
\]
so that $\gamma \mapsto Q(\gamma, \bar{p})$ is a Krein-unitary representation of the subgroup $G_{\bar{p}}$
of $SL(2, \mathbb{C})$ stationary for $\bar{p}$. Thus the initial representation is equivalent to the representation
(we denote it by the same letters $U, T$ as the initial one) whose action on the decomposition functions is given by the following formula
\begin{multline}\label{U,T}
U(\alpha)\widetilde{\psi}(p) = Q(\gamma(\alpha,p),\bar{p}) \widetilde{\psi} (\Lambda(\alpha)p), \\
T(a) \widetilde{\psi}(p) = e^{i a \cdot p} \widetilde{\psi}(p) =  e^{i g_{M}(a, p)} \widetilde{\psi}(p), 
\end{multline} 
where $\gamma \mapsto Q(\gamma, \bar{p})$ is a Krein-unitary representation of 
the subgroup $G_{\bar{p}}$ stationary for a fixed point $\bar{p}$ belonging to the orbit 
$\mathscr{O}_{\bar{p}} = {}$ sp$(P^0 , \ldots , P^3)$.

Our next step is to find an explicit formula for the unitary and Krein-unitary transformation
(we denote it likewise by $W$) which applied to vector states $\widetilde{\psi}$ of the representation
$U_{(a,1)} = T(a)$, $U_{(0,\alpha)} = U(\alpha)$ with $T(a)$ and $U(\alpha)$ given by (\ref{U,T})
gives a transformation formula with a multiplier independent of $p \in \mathscr{O}_{\bar{p}}$, i . e.
$W$ is such that the Fourier transform 
\begin{equation}\label{F(varphi)}
\varphi (x) = (2\pi)^{-3/2} \int \limits_{\mathscr{O}_{\bar{p}}} \widetilde{\varphi}(p) e^{-ip \cdot x} \, 
\ud \mu |_{{}_{\mathscr{O}_{\bar{p}}}} (p)
\end{equation}
of $\widetilde{\varphi} = W\widetilde{\psi}$ has a local transformation law, where $\ud \mu (p)$
is the invariant measure induced on the orbit $\mathscr{O}_{\bar{p}}$ by the Lebesgue measure 
on $\mathbb{R}^4$.

   To this end we need a representation $\alpha \mapsto V(\alpha)$ of $SL(2, \mathbb{C})$ 
acting in the Krein space $(\mathcal{H}_{\bar{p}}, \mathfrak{J}_{\bar{p}})$ which extends 
the Krein-unitary representation $\gamma \mapsto Q(\gamma, \bar{p})$ of the subgroup
$G_{\bar{p}} \subset SL(2, \mathbb{C})$ to a representation of the whole $SL(2, \mathbb{C})$
group. $V$ need not be Krein-unitary (resp. unitary in case $\mathfrak{J}_{\bar{p}} = \bold{1}$) 
It turns out that such extensions $V$
do exist for the Krein-unitary (resp. unitary in case $\mathfrak{J}_{\bar{p}} = \bold{1}$) 
representations  $\gamma \mapsto Q(\gamma, \bar{p})$ associated with the 
representations concentrated on single orbits which arise in the process of decomposition of the 
tensor products of representations acting in single particle subspaces. For example they are well known for the 
representations $\gamma \mapsto Q(\gamma, \bar{p})$ associated with the representations concentrated on 
single orbits which arise in decompositions of tensor
products of spin one-half, non-zero-mass representations (in this case  $\mathfrak{J}_{\bar{p}} = \bold{1}$
and the representations $\gamma \mapsto Q(\gamma, \bar{p})$ are unitary.

Namely let us define the transformation $W$ whose action on decomposition functions is defined 
in the following manner
\begin{equation}\label{W2}
\widetilde{\varphi}(p) = W \widetilde{\psi}(p) = V(\beta(p)^{-1})\widetilde{\psi}(p).
\end{equation}
Then we have
\begin{multline*}
W U(\alpha)W^{-1}\widetilde{\varphi}(p) = V \big( \beta(p)^{-1} \big) V\big( \gamma(\alpha,p) \big) 
V \big(\beta(\Lambda(\alpha)p) \big) \widetilde{\varphi}(\Lambda(\alpha)p) \\
= V \Big( \beta(p)^{-1} \, \beta(p) \, \alpha \, \beta(\Lambda(\alpha)p)^{-1} \, \beta(\Lambda(\alpha)p) \Big) \widetilde{\varphi}(\Lambda(\alpha)p) = V(\alpha) \widetilde{\varphi}(\Lambda(\alpha)p),
\end{multline*}
therefore
\begin{equation}\label{wuw^-1}
W U(\alpha)W^{-1}\widetilde{\varphi}(p) = V(\alpha) \widetilde{\varphi}(\Lambda(\alpha)p),
\end{equation}
such that the Fourier transform $\varphi$ defined by (\ref{F(varphi)})  of $\widetilde{\varphi}$
has a local transformation formula
\begin{multline}\label{loc.tr.x}
U(\alpha) \varphi(x) = V(\alpha) \varphi (x \Lambda(\alpha^{-1})) 
= V(\alpha) \varphi (x_\nu {\Lambda(\alpha^{-1})_{\mu}}^\nu) 
\end{multline}
(summation with respect to $\nu$), where we have used again the symbol $U$ for the representation in the space of Fourier transforms $\varphi$ hoping that it will not cause any serious misunderstandings.

Let
\[
\Big( \mathcal{H} = \int \limits_{sp(P^0 , \ldots , P^3) \cong \mathscr{O}_{\bar{p}}} \mathcal{H}_{\bar{p}} 
\, \ud \mu |_{{}_{\mathscr{O}_{\bar{p}}}} (p), \,\,\,
\mathfrak{J} =  \int \limits_{sp(P^0 , \ldots , P^3) \cong \mathscr{O}_{\bar{p}}} \mathfrak{J}_{\bar{p}} 
\, \ud \mu |_{{}_{\mathscr{O}_{\bar{p}}}} (p) \Big)
\]
be the Krein space of the representation (\ref{U,T}), which we may assume to 
be equal to the Krein space of the initial representation, as the transformation $W$ given by
(\ref{W1}) preserves te core set $\mathfrak{D}$ of the initial representation and is Krein-isometric.

Let 
\[
\Big( \mathcal{H'} = \int \limits_{sp(P^0 , \ldots , P^3) \cong \mathscr{O}_{\bar{p}}} \mathcal{H'}_{p} 
\, \ud \mu |_{{}_{\mathscr{O}_{\bar{p}}}} (p), \,\,\,
\mathfrak{J'} =  \int \limits_{sp(P^0 , \ldots , P^3) \cong \mathscr{O}_{\bar{p}}} \mathfrak{J'}_{p} 
\, \ud \mu |_{{}_{\mathscr{O}_{\bar{p}}}} (p) \Big),
\]
be the Krein space with the Hilbert space inner product in $\mathcal{H'}$ defined by
\[
(\widetilde{\varphi}, \widetilde{\varphi}') 
= \int \limits_{sp(P^0 , \ldots , P^3) \cong \mathscr{O}_{\bar{p}}} 
\Big( \widetilde{\varphi}(p), \widetilde{\varphi}'(p) \Big)_{p} 
\, \ud \mu |_{{}_{\mathscr{O}_{\bar{p}}}} (p), 
\]
where 
\[
\Big( \widetilde{\varphi}(p), \widetilde{\varphi}'(p) \Big)_{p} = 
\Big(  \widetilde{\varphi}(p), 
V(\beta(p))^* V(\beta(p)) \widetilde{\varphi}'(p)  \Big)_{\mathcal{H}_{\bar{p}}}, \,\,\,\,
\widetilde{\varphi}(p), \widetilde{\varphi}'(p) \in \mathcal{H'}_{p},
\]
with the inner product $\big( \cdot , \cdot  \big)_{\mathcal{H}_{\bar{p}}}$ of the
Hilbert space $\mathcal{H}_{\bar{p}}$ (with the convention assumed here, that it is 
conjugate linear in the first variable\footnote{This convention is assumed in most of the physical literature.});
and let the decomposition components of the fundamental symmetry $\mathfrak{J'}$ defined by
\[
\mathfrak{J'}_{p} = V(\beta(p)^{-1}) \, \mathfrak{J}_{\bar{p}} \, V(\beta(p)).
\]
We then have the following   

\begin{twr*}
The transformation $W$ defined by (\ref{W2}), which transforms $\widetilde{\psi}$ belonging to the 
Krein space $(\mathcal{H}, \mathfrak{J})$
of the initial representation, equal to the Krein space of the representation defined by (\ref{U,T}),
onto the  the Krein space $(\mathcal{H'}, \mathfrak{J'})$ of elements $\widetilde{\varphi}$ of the representation 
(\ref{wuw^-1}), is unitary and Krein-unitary. 
\end{twr*}

\begin{rem*}
The components $\mathfrak{J'}_{p}$ of the decomposition of the fundamental symmetry $\mathfrak{J'}$
depend in general on $p \in \mathscr{O}_{\bar{p}}$, because $V(\beta(p))$ -- although being 
Krein-unitary and unitary in the respective Krein space $\mathcal{H'}_{p}, \mathfrak{J'}_{p}$ for all $p$ -- are not in general unitary in the Hilbert space $\mathcal{H}_{\bar{p}}$:
\begin{multline*}
\mathfrak{J'}_p = V(\beta(p)^{-1}) \, \mathfrak{J}_{\bar{p}} V(\beta(p)) \\
= \mathfrak{J}_{\bar{p}} V(\beta(p))^* \, \mathfrak{J}_{\bar{p}} \mathfrak{J}_{\bar{p}} V(\beta(p))
= \mathfrak{J}_{\bar{p}} V(\beta(p))^* \, V(\beta(p)), 
\end{multline*}
where $\mathfrak{J}_{\bar{p}}$ does not depend on $p$ and $V(\beta(p))^* V(\beta(p))$ depends on $p$.
\end{rem*}

Therefore construction of a local single particle wave function is nothing more than the the application of the above construction of the transformation $\widetilde{\psi}
\mapsto \widetilde{\varphi}$ with the properties indicated by the above theorem. The crucial point being that
the Fourier transform (\ref{F(varphi)}) of the transformed $\widetilde{\varphi}$ has a local transformation law.

\vspace*{0.5cm}

{\bf ACKNOWLEDGEMENTS}

\vspace*{0.3cm}

The author is indebted for helpful discussions to prof. A. Staruszkiewicz.

\end{document}